\documentclass[floatfix,reprint,prl,aps,twocolumn,superscriptaddress,longbibliography]{revtex4-2}
\usepackage[bitstream-charter]{mathdesign}
\usepackage{float}
\usepackage{amsmath}
\DeclareMathOperator{\sign}{sgn}
\usepackage{graphicx}
\usepackage{dcolumn}
\usepackage{bm}
\usepackage{calrsfs}
 
\usepackage[colorlinks=true,linkcolor=blue,urlcolor=blue,citecolor=blue,pdfusetitle]{hyperref}

\begin{document}

\title{Bardeen-Cooper-Schrieffer State Representation and Pairing in the Fermionic Tonks-Girardeau Gas}

\author{Francesc Sabater}
\affiliation{Departament de F{\'i}sica Qu{\`a}ntica i Astrof{\'i}sica, Facultat de F{\'i}sica, Universitat de Barcelona, E-08028 Barcelona, Spain}
\affiliation{Institut de Ci{\`e}ncies del Cosmos, Universitat de Barcelona, ICCUB, Mart{\'i} i Franqu{\`e}s 1, E-08028 Barcelona, Spain}
\author{Abel Rojo-Francàs}
\affiliation{Departament de F{\'i}sica Qu{\`a}ntica i Astrof{\'i}sica, Facultat de F{\'i}sica, Universitat de Barcelona, E-08028 Barcelona, Spain}
\affiliation{Institut de Ci{\`e}ncies del Cosmos, Universitat de Barcelona, ICCUB, Mart{\'i} i Franqu{\`e}s 1, E-08028 Barcelona, Spain}
\author{Grigori E. Astrakharchik}
\affiliation{Departament de F\'isica, Universitat Polit\`ecnica de Catalunya, 
Campus Nord B4-B5, E-08034 Barcelona, Spain}

\author{Bruno Juli{\'a}-D{\'i}az}
\affiliation{Departament de F{\'i}sica Qu{\`a}ntica i Astrof{\'i}sica, Facultat de F{\'i}sica, Universitat de Barcelona, E-08028 Barcelona, Spain}
\affiliation{Institut de Ci{\`e}ncies del Cosmos, Universitat de Barcelona, ICCUB, Mart{\'i} i Franqu{\`e}s 1, E-08028 Barcelona, Spain}

\date{\today}

\begin{abstract}
 We demonstrate that the ground state of the fermionic Tonks-Girardeau gas is a number-conserving Bardeen-Cooper-Schrieffer (BCS) state. This offers a compelling new perspective on describing the ground state wave function of the fermionic Tonks-Girardeau gas, enabling a complete description in second quantization. The proposed wave function is constructed based on the occupation numbers and natural orbitals of the one-body density matrix and is valid under any external potential. Specifically, we provide the exact  coefficients that allow to describe the fermionic Tonks-Girardeau gas as a number-conserving BCS state. Furthermore, the proposed wave function's form in second quantization allows us to derive the necessary expectation values to detect pairing in the fermionic Tonks-Girardeau gas. Through this approach, we demonstrate and show how to detect that the fermionic Tonks-Girardeau gas not only exhibits non-trivial quantum correlations but is also a paired state.
\end{abstract}
                             
\maketitle
 \textbf{Introduction.} 
 One-dimensional fermionic and bosonic systems are of particular interest 
 due to their pronounced quantum effects and mathematical tractability, which aid in the development and validation of theoretical models~\cite{sutherland2004beautiful}. Ultracold gases serve as a platform for studying one-dimensional quantum many-body systems with short-range interactions~\cite{Lewenstein_2007}. Such one-dimensional systems have already been experimentally realized using confinement in quasi-1D (highly elongated) harmonic traps~\cite{Greiner2001,Gorlitz2001,HallerNägerl09,KaoLev21,Zurn2012,Wenz_2013}. At the same time, the interaction between particles can be precisely controlled and tuned~\cite{GrangerBlume}.

 A characteristic feature of one-dimensional systems is the mapping between fermionic and bosonic systems. This mapping allows to describe strongly correlated bosonic (fermionic) systems as the symmetrized (antisymmetrized) version of simpler fermionic (bosonic) systems. This fermion-boson duality emerges from the Girardeau mapping, for which the bosonic and fermionic systems share identical diagonal properties~\cite{girardeau_mapping, Cheon1999, Sekino}. A prominent example is the Tonks-Girardeau gas, describing one-dimensional, strongly repulsive, hard-core bosons mapped onto a system of ideal fermions under the same external potential~\cite{girardeau_mapping, Girardeau2001, Cheon1999, Sekino, linieger, paredes2004tonks, Marmorini_2016,Yukalov_2005}. The generalization of this mapping to arbitrary interaction strengths, by Cheon and Shigehara \cite{Cheon1999}, allows for mapping the ground state of strongly attracting $p$-wave fermions to the ground state of noninteracting bosons. This system is referred to as the fermionic Tonks-Girardeau (FTG) gas. Its ground state wave function (WF) is
\begin{equation}
\label{gs attractive fermions}
\psi_F(x_1,\dots ,x_N)=\psi_B(x_1,\dots ,x_N)\prod\limits_{j<k}^N \sign{(x_j-x_k)},
\end{equation}
and is related to that of an ideal Bose gas, $\psi_B(x_1,\dots,x_N)=\prod_{i=1}^{N}\phi(x_i)$, where $\phi(x)$ is the single-particle ground state~\cite{GirardeauNguyenOlshanii2004,GirardeauMinguzzi2006,MinguzziGirardeau2006,BenderErkerGranger2005,Kościk_2020,koscik,delcampo2006,Hao2007}.

In a recent work, we computed the one-body density matrix (OBDM) 
$\rho_1({x},{x'})=\langle\hat \Psi^\dagger({x})\hat\Psi({x'})\rangle$ for 
the FTG gas and diagonalized it~\cite{Sabater}. 
We presented analytical expressions for the eigenvalues of the OBDM, $\lambda_{k}^{(N)}$, i.e. the occupation numbers. We showed that the occupation numbers are always doubly degenerate, and presented expressions for the corresponding degenerate natural orbitals, $\chi_{k+}(x)$ and $\chi_{k-}(x)$.

With these expressions, we showed that the total density of the FTG gas, which is equivalent to the density of $N$ non-interacting bosons,
\begin{equation}
n(x) = N|\phi(x)|^2,
\end{equation}
can be explained by the formation of $N/2$ pairs of fermions in the orbitals $\chi_{k+}(x)$ and $\chi_{k-}(x)$ with a density of $2|\phi(x)|^2$ per pair in the even case. In the odd case, we reasoned that $(N-1)/2$ pairs are formed, with the remaining fermion occupying the single-particle ground state~\cite{Sabater}.\\

In this Letter, we show that the FTG gas is a number-conserving BCS state, implying that the FTG gas can be fully described and characterized by the same type of states appearing in the BCS theory of superconductivity~\cite{BCS}. This provides a new way to express the ground state WF of the FTG gas, as given by Eq.~(\ref{gs attractive fermions}), explicitly containing the observed pairing in Ref.~\cite{Sabater}. We provide a detailed proof that the proposed WF is equivalent to the pair-product one~(\ref{gs attractive fermions}). Remarkably, we derive the proposed ground state WF solely from information obtained from the OBDM, that is, its occupation numbers, natural orbitals, and our understanding of the pairing mechanism. This is noteworthy because in most systems recovering the WF from OBDM information alone is neither straightforward nor possible. The proposed WF is more suitable for working within a second quantization framework than the pair product form~(\ref{gs attractive fermions}). It enables the calculation of important expectation values involving creation and annihilation operators, paving the way for new research directions on the FTG gas that were not possible with the WF~(\ref{gs attractive fermions}). Lastly, we provide the expectation values of a set of measurable operators, which show that the FTG gas is a paired state within the framework for pairing derived in Ref.~\cite{pairing_krauss2009}.

\textbf{A novel expression for the FTG gas ground state.} For clarity, we initially present the simpler case of $N=2$. From Ref.~\cite{Sabater}, we know that when dealing only with two fermions, these two fermions form a pair occupying the orbitals $\chi_{k+}(x)$ and $\chi_{k-}(x)$ for a given $k$ with probability $\lambda_k^{(2)} = 8/[\pi(2k-1)]^2$. Based on this, we postulate that the ground state of the two-particle FTG gas can be written as:
\begin{equation}
\label{wf 2 particles}
\psi_T(x_1,x_2)=\sum_{k=1} \sqrt{\lambda_k^{(2)}} \varphi_k(x_1,x_2),
\end{equation}
where $\varphi_k(x_1,x_2)$ is the Slater determinant formed by the states $\chi_{k+}(x)$ and $\chi_{k-}(x)$. We denote our proposed WF as $\psi_T$ to distinguish it from the well-known FTG gas ground state $\psi_F$ given by Eq.~(\ref{gs attractive fermions}). The proposed WF is normalized since:
\begin{equation}
    \int\int |\psi_T(x_1,x_2)|^2 dx_1dx_2=\sum_{k=1} \lambda_k^{(2)} = N/2 = 1.
\end{equation}
The proof that $\psi_T = \psi_F$ is straightforward and included here. For larger numbers of particles, we provide the proof in the Supplementary Material~\cite{supplemental}.
We start by noting that 
\begin{equation}
    \varphi_k(x_1,x_2)\!=\!\phi(x_1)\phi(x_2)\sqrt{2}\sin\left[(2k-1)\pi(y_1-y_2)\right].
\end{equation}
where $y_i=F(x_i)=\int_{-\infty}^{x_i}|\phi(z)|^2  dz$. Then,
\begin{equation}
\begin{aligned}
    \frac{\psi_T(x_1,x_2)}{\phi(x_1)\phi(x_2)} &= \sum_{k=1}\sqrt{\lambda_k^{(2)}}\sqrt{2}\sin\left[(2k-1)\pi(y_1-y_2)\right] \\
    &= \frac{4}{\pi}\sum_{k=1}\frac{\sin\left[(2k-1)\pi(y_1-y_2)\right]}{2k-1} \\
    &=\sign(y_1-y_2) =\sign[F(x_1)-F(x_2)], 
\end{aligned}
\end{equation}
for $F(x_1)-F(x_2)\in(-1,1)$. Since $F(x)$ is a monotonically increasing function, $\sign[F(x_1)-F(x_2)]=\sign(x_1-x_2)$. Therefore,
\begin{equation}
   \psi_T(x_1,x_2)=\phi(x_1)\phi(x_2)\sign(x_1-x_2)=\psi_F(x_1,x_2),
\end{equation}
as we wanted to prove. Remarkably, Eq.~(\ref{wf 2 particles}) can be interpreted as the Slater decomposition of the FTG gas. The Slater decomposition can be understood as analogous to the Schmidt decomposition for fermionic, thus indistinguishable, systems~\cite{correlation2fermions,Eckert_2002}. Since the Slater decomposition obtained contains more than one non-zero coefficient $\lambda_k^{(2)}$, the FTG gas has a Slater rank greater than one. This implies that the FTG gas exhibits non-trivial quantum correlations according to Refs.~\cite{correlation2fermions,Eckert_2002}.

The proposed WF for the $N=3$ case follows from a similar reasoning to the $N=2$ case. As shown in Ref.~\cite{Sabater}, only one pair is formed in the states $\chi_{k+}(x)$ and $\chi_{k-}(x)$ while the third fermion occupies the single-particle ground state. Based on this, we postulate 
\begin{equation}
    \psi_T(x_1,x_2,x_3)=\sum_{k=1} \sqrt{\lambda_k^{(3)}} \varphi_{0,k}(x_1,x_2,x_3),
\end{equation}
where $\varphi_{0,k}(x_1,x_2,x_3)$ is the Slater determinant formed by the states $\chi_{k+}(x)$, $\chi_{k-}(x)$, and $\phi(x)$. Again, the WF is properly normalized \cite{Sabater}:
\begin{equation}
    \sum_{k=1} \lambda_k^{(3)} = \frac{(N-1)}{2} = 1.
\end{equation}
Note that the non-degenerate eigenvalue $\lambda_0^{(3)}=1$ is not included in the summation. In the Supplementary Material, we provide a proof that $\psi_T(x_1,x_2,x_3) = \psi_F(x_1,x_2,x_3)$~\cite{supplemental}.

The next step is to obtain the WF for $N=4$ fermions. In this case, there are two different pairs at two different $k$'s, $k_1$ and $k_2$. Based on that, we propose 
\begin{equation}  \psi_T(x_1,\ldots,x_4)=\sum_{k_1<k_2} \sqrt{p_4(k_1,k_2)} \varphi_{k_1,k_2}(x_1,\ldots,x_4),
\end{equation}
where $\varphi_{k_1,k_2}(x_1,\ldots,x_4)$ is the Slater determinant containing the 4 states involved in the two pairs $k_1$ and $k_2$. We denote by $p_4(k_1,k_2)$ the coefficients that represent the probability of having pairs $k_1$ and $k_2$ occupied. 
Initially, one might consider $p_4(k_1,k_2)=\lambda_{k_1}^{(4)}\lambda_{k_2}^{(4)}$, but these coefficients imply improper normalization. This is because we are treating the probabilities of filling each pair as independent when they are not. The probability of filling the second pair $k_2$ depends on which pair, $k_1$, is filled since, to fulfill the Pauli exclusion principle, we can not fill again $k_1$. Therefore, to obtain the proper coefficients $p_4(k_1,k_2)$, we should account for the dependencies between the probabilities of filling each pair.

Following the same reasoning regarding the formation of pairs, we find that the WF in the case of $N$ fermions (here $N$ is even) is
\begin{equation}
\label{wf N even}
    \!\!\psi_T(\!x_1,\!\ldots \!,x_N\!)\!=\!\!\!\!\!\!\!\!\!\!\sum_{k_1<\ldots\!<k_{N_p}} \!\!\!\!\!\!\!\!\!\sqrt{p_N(k_1,\!\ldots\!,k_{N_p})} \varphi_{k_1,\ldots \!,k_{N_p}}\!\!(\!x_1,\!\ldots\!,x_N\!),\!\!
\end{equation}
where $N_P=N/2$ denotes the number of pairs and $\varphi_{k_1,\ldots ,k_{N_P}}(x_1,\ldots,x_N)$ is the Slater determinant containing the $N$ states $\chi_{k_1+}(x)$, $\chi_{k_1-}(x)$, \ldots, $\chi_{k_{N_P}+}(x)$, $\chi_{k_{N_P}-}(x)$. 
The expression for odd number of fermions is similar, but it includes an unpaired fermion in the single-particle ground state,
\begin{equation}
\label{wf N odd}
   \!\!\!\!\psi_T(\!x_1,\!\ldots \!,\!x_N\!)\!=\!\!\!\!\!\!\!\!\!\!\sum_{k_1<\ldots\!<k_{N_p}} \!\!\!\!\!\!\!\!\!\sqrt{p_N(k_1,\!\ldots\!,\!k_{N_p})} \varphi_{0,k_1,\ldots ,k_{N_p}}\!\!(\!x_1,\!\ldots\!,\!x_N\!),\!\!
\end{equation}
where now $N_P=(N-1)/2$ and $\varphi_{0,k_1,\ldots ,k_{N_P}}(x_1,\ldots,x_N)$ is the Slater determinant containing the $N$ states $\chi_{k_1+}(x)$, $\chi_{k_1-}(x)$, \ldots, $\chi_{k_{N_P}+}(x)$, $\chi_{k_{N_P}-}(x)$, and importantly, $\phi(x)$.

Inspired by the fact that the probability of filling a pair $k$ cannot be treated independently from the other filled pairs, we postulate the following recursive formula for $p_N(k_1,\ldots,k_{N_P})$,
\begin{equation}
   \label{pN recursive} \!\!\!p_N(k_1,\!\ldots\!,k_{N_p})=\lambda_{k_1}^{(N)}\frac{p_{N-2}(k_2,\ldots,k_{N_p})}{1-\!\!\sum_{(l<i<\!\ldots\!)\neq k_1}\!p_{N-2}(k_1,l,i,\!\ldots\!)}.
\end{equation}
This formula can be understood as a Bayes conditioned probability $P(A,B)=P(A)P(B|A)$. Where in our case $A$ is the probability of filling the pair $k_1$. Here, $P(B|A)$ is the probability of filling the $N_P-1$ remaining pairs conditioned on $k_1$ being already filled. Importantly, we can check that if $p_{N-2}$ is properly normalized (which we know for $N=2$ and $N=3$), then $p_N$ is also properly normalized. The short demonstration of proper normalization starts with
\begin{equation}
    \!\!\sum_{k_1<\ldots<k_{N_p}} \!\!\!p_N(k_1, \ldots, k_{N_p}) 
    = \frac{1}{N_p!}\! \sum_{k_1 \neq \ldots \neq k_{N_p}}\!\! p_N(k_1, \ldots, k_{N_p}) 
\end{equation}
which equals
\begin{equation}
    \!\!\frac{(N_p\!\!-\!1)!}{N_p!} \! \!\sum_{k_1} \lambda_{k_1}^{(N)} \! \! \!\!\!\!\!   \sum_{(k_2 < \!\ldots\! < k_{N_p}) \neq k_1}  \! \! \!\!\!\frac{p_{N-2}(k_2, \ldots, k_{N_p})}{1 -  \! \!\sum_{(l< \!\ldots\!) \neq k_1}  \!  p_{N-2}(k_1, l, \!\ldots\!)}. 
\end{equation}
The second summation $\sum_{(k_2 < \ldots < k_{N_P}) \neq k_1}$ equals unity. Finally,
\begin{equation}
    \sum_{k_1<\ldots<k_{N_P}} p_N(k_1, \ldots, k_{N_P}) = \frac{1}{N_P} \sum_{k_1} \lambda_{k_1}^{(N)} = 1,
\end{equation}
as we wanted to prove. Both Eq.~(\ref{pN recursive}) and the proof of proper normalization are valid for the even and odd cases. From the recursive formula Eq.~(\ref{pN recursive}), we can derive an explicit formula for $p_N(k_1, \ldots, k_{N_P})$. First,
\begin{equation}
    \sum_{(l<i<\ldots)\neq k_1} p_{N-2}(k_1, l, i, \ldots) = \lambda_{k_1}^{(N-2)},
\end{equation}
which can be easily derived using Eq.~(\ref{pN recursive}) since the summation does not involve $k_1$. As a consequence, Eq.~(\ref{pN recursive}) is simplified to 
\begin{equation}
   \label{pN recursive explicit} p_N(k_1,\ldots,k_{N_p})=\lambda_{k_1}^{(N)}\frac{p_{N-2}(k_2,\ldots,k_{N_p})}{1-\lambda_{k_1}^{(N-2)}}.
\end{equation}
Then, it follows from expressing $\lambda_{k_1}^{(N)}$ as a function of $\lambda_{k_1}^{(N-2)}$, using the recursive formula for the occupation numbers derived in Ref.~\cite{Sabater}, that
\begin{equation}
    p_N(k_1, \ldots, k_{N_P}) = \frac{N(N-1)}{(k_1-1/2)^2\pi^2} p_{N-2}(k_2, \ldots, k_{N_P}),
\end{equation}
for the even case, and
\begin{equation}
    p_N(k_1, \ldots, k_{N_P}) = \frac{N(N-1)}{k_1^2\pi^2} p_{N-2}(k_2, \ldots, k_{N_P}),
\end{equation}
for the odd case. Finally, by starting from $p_2(k)=\lambda_k^{(2)}$ and $p_3(k)=\lambda_k^{(3)}$, we arrive at the final explicit formulas,
\begin{equation}
\label{pN combined}
p_N(k_1, \ldots, k_{N_P}) =
\left\{\begin{array}{ll}
\frac{N!}{\pi^N} \prod_{i=1}^{N_P} \frac{1}{(k_i-1/2)^2}, & \text{if } N \text{ is even} \\
\frac{N!}{\pi^{(N-1)}} \prod_{i=1}^{N_P} \frac{1}{k_i^2}, & \text{if } N \text{ is odd.}
\end{array}\right.
\end{equation}
This form of $p_N$ fulfills normalization, as already shown. Additionally, it is invariant under the chosen order of $k$'s. While these two properties are necessary for a suitable form of $p_N$, they do not guarantee that our guess is the correct one. In the Supplementary Material, we provide a detailed proof that the proposed WF $\psi_T$~(\ref{wf N even},~\ref{wf N odd}) along with the proposed form of $p_N$~(\ref{pN combined}) satisfies $\psi_T=\psi_F$~\cite{supplemental}. In addition to the mathematical proof, we performed several numerical checks on the equivalency of the two WF's. 

Our proposed WF, $\psi_T$, is mathematically equivalent to the original WF of the FTG gas, $\psi_F$~(\ref{gs attractive fermions}). This equivalence demonstrates that $\psi_T$, which is constructed solely based on the formation of pairs without including higher-order correlation terms such as three-body correlations, fully describes the FTG gas. The ability to describe the FTG gas state entirely through pair formation is particularly noteworthy, as previous studies have shown that three-body properties in spinless $p$-wave fermions are non-negligible~\cite{three_body_1,three_body_2,three_body_3}.

\textbf{The FTG gas as a BCS state.}
In this section, we demonstrate that the WF of the FTG gas in second quantization form can be expressed as a number-conserving BCS state. The paired states appearing in the BCS theory of superconductivity~\cite{BCS} can be described as a superposition of number-conserving BCS states $|\psi_{BCS}^{(N)}\rangle$,
\begin{equation}
    |\psi_{BCS}\rangle=\sum_N \beta_N |\psi_{BCS}^{(N)}\rangle.
\end{equation}
The specific form of $|\psi_{BCS}^{(N)}\rangle$ is~\cite{BCS}
\begin{equation}
    \label{bcs state 1}|\psi_{BCS}^{(N)}\rangle=C_N\left(\sum_k\alpha_k P_k^\dagger\right)^{N_P}|0\rangle,
\end{equation}
where $C_N$ is a normalization constant, $P_k^\dagger=a_k^\dagger a_{-k}^\dagger$, and $|0\rangle$ is the vacuum state. Here, $a_k^\dagger$ is the creation operator that creates a fermion in the single-particle orbital labeled by $k$. In our system, this means $a_k^\dagger$ creates a fermion in the orbital $\chi_{k+}(x)$, while $a_{-k}^\dagger$ creates one in $\chi_{k-}(x)$.

In the state $|\psi_{BCS}^{(N)}\rangle$, all $N_P$ pairs occupy the same two-particle state given by $\sum_k \alpha_k P_k^\dagger$. The specific case where the single-particle orbitals are spin-momentum states, i.e., $k = (\vec{k}, \uparrow)$ and $-k = (-\vec{k}, \downarrow)$ corresponds to the standard BCS theory of superconductivity.
Equation~(\ref{bcs state 1}) can be alternatively written as \cite{pairing_krauss2009}
\begin{equation}
    \label{bcs state 2}
    |\psi_{BCS}^{(N)}\rangle=C_N N_p! \sum_{k_1<\ldots<k_{N_p}} \alpha_{k_1}\!\ldots\alpha_{k_{N_p}} P_{k_1}^\dagger \!\ldots P_{k_{N_p}}^\dagger |0\rangle.
\end{equation}
Since our new formulation of the ground state WF is a superposition of Slater determinants, it is particularly convenient to express it in second quantization. For the even case, we have,
\begin{equation}
\label{wf N even 2n quan}
    |\psi_T\rangle = \sum_{k_1<\ldots<k_{N_p}} \sqrt{p_N(k_1,\ldots,k_{N_p})} P_{k_1}^\dagger \!\ldots P_{k_{N_p}}^\dagger |0\rangle,
\end{equation}
where  $P_{k}^\dagger = a_{k+}^\dagger a_{k-}^\dagger$. Since $p_N$~(\ref{pN combined}) factorizes in all the $k_i$'s, we can rewrite our state as, 
\begin{equation}
    |\psi_T\rangle = \sum_{k_1<\ldots<k_{N_p}} \alpha_{k_1} \!\ldots \alpha_{k_{N_p}} P_{k_1}^\dagger \!\ldots P_{k_{N_p}}^\dagger |0\rangle,
\end{equation}
with 
\begin{equation}
\label{alpha even}
\alpha_{k_i} = \frac{N!^{1/N}}{\pi} \frac{1}{(k_i - 1/2)}.
\end{equation}
Thus, we have clearly expressed the ground state WF of the even FTG gas as a number-conserving BCS state and provided the explicit expressions for the coefficients $\alpha_{k_i}$. The normalization constant $C_N$ is found to be $C_N = 1/N_p!$. To express the odd case WF in second quantization, we need to add the creation operator for a single particle in the single-particle ground state,
\begin{equation}
\label{wf N odd 2n quan}
    |\psi_T\rangle = a_0^\dagger \sum_{k_1<\ldots<k_{N_p}} \sqrt{p_N(k_1,\ldots,k_{N_p})} P_{k_1}^\dagger \!\ldots P_{k_{N_p}}^\dagger |0\rangle.
\end{equation}
Again, we express $p_N$~(\ref{pN combined}) as a product of $\alpha_{k_i}$'s, getting to,
\begin{equation}
    |\psi_T\rangle = a_0^\dagger  \sum_{k_1<\ldots<k_{N_p}} \alpha_{k_1} \!\ldots \alpha_{k_{N_p}} P_{k_1}^\dagger \!\ldots P_{k_{N_p}}^\dagger |0\rangle,
\end{equation}
where now
\begin{equation}
\label{alpha_odd}
    \alpha_{k_i} = \frac{N!^{1/(N-1)}}{\pi} \frac{1}{k_i}.
\end{equation}
Thus, we have expressed the ground state WF of the odd FTG gas as a number-conserving BCS state, with the addition of a particle in the single-particle ground state.
Therefore, we have seen that starting from the general form of a BCS state, we can fully characterize and describe the FTG ground state WF. The demostration is based on the novel form of the ground state WF and the fact that $p_N$ factorizes in the $k_i$'s.

\textbf{Detecting pairing in the FTG gas.} The pairing phenomenon can be treated as a quantum correlation and can be formally defined within quantum information theory~\cite{pairing_krauss2009}. A fermionic state is said to be paired if there exists a set of operators containing at most two creation and two annihilation operators such that their expectation values cannot be reproduced by any separable state~\cite{pairing_krauss2009}. Following this definition, it can be formally proved that any number-conserving BCS state is a paired state~\cite{pairing_krauss2009}. The proof is based on the set of operators
\begin{equation}
\vec{O}_3 = 
\begin{pmatrix}
n_k + n_{-k} + n_l + n_{-l} \\
n_k n_{-k} + n_l n_{-l} \\
a_k^\dagger a_{-k}^\dagger a_{-l} a_l + \text{h.c.}
\end{pmatrix}.
\end{equation}
The expectation values of $\vec{O}_3$ obtained with a number-conserving BCS state can not be reproduced by any separable state ~\cite{pairing_krauss2009}. Specifically, the expectation value of 
\begin{equation}
    \begin{aligned}
        H_1 = & \frac{1}{2}(n_k + n_{-k} + n_l + n_{-l}) - (n_k n_{-k} + n_l n_{-l}) \\
              & -(a_k^\dagger a_{-k}^\dagger a_{-l} a_l + \text{h.c.}),
    \end{aligned}
\end{equation}
is greater or equal than zero, $\langle H_1 \rangle \geq 0$,  for any separable state. For a number-conserving BCS state, $\langle H_1 \rangle < 0$ \cite{pairing_krauss2009}.

The set of operators $\vec{O}_3$ does not include any creation or annihilation operator acting on the ground state $k=0$. Therefore, the proof can be immediately extended to the even-number FTG gas and also to the odd-number FTG gas, since the addition of a particle in the single-particle ground state does not alter the expectation values of $\vec{O}_3$. Therefore, using the newly proposed shape of the WF, we conclude that the FTG gas forms a paired state.

Making use of our proposed ground state WF we are also able to compute the values of $\vec{O}_3$ for the FTG gas, 
\begin{equation}
\label{O3 FTG}
\langle\vec{O}_3^{FTG}\rangle \!\!=\! 
\begin{pmatrix}
2\left(\lambda_k^{(N)}+\lambda_l^{(N)}\right) \\
\lambda_k^{(N)}+\lambda_l^{(N)}\\
2\!\sum_{(k_2 < \ldots < k_{N_P}) \neq l,k} \!\!\sqrt{\!p(l, k_2 \ldots)p(k, k_2 \ldots)}
\end{pmatrix}.
\end{equation}
These values imply
\begin{equation}
    \langle H_1\rangle = -2\!\sum_{(k_2 < \ldots < k_{N_P}) \neq l,k} \!\!\sqrt{\!p(l, k_2 \ldots)p(k, k_2 \ldots)}<0.
\end{equation}
The detailed calculations of these expectation values are provided in the Supplementary Material~\cite{supplemental}. The calculations are relatively straightforward due to the suitable form in second quantization of the proposed WF. The specific values of $\langle\vec{O}_3^{FTG}\rangle$ are of interest because they are the main quantities to measure in order to detect pairing in a fermionic system. Therefore, if $\langle\vec{O}_3^{FTG}\rangle$ were to be measured, one could observe the existence of pairing in the FTG gas and determine that the prepared system is indeed a FTG gas.

\textbf{Conclusions.} 
In this Letter, we present a new expression for the ground state wave function of the fermionic Tonks-Girardeau gas, as given by Eqs.~(\ref{wf N even}) and~(\ref{wf N odd}) for even and odd number of fermions, respectively. This wave function is based exclusively on the occupation numbers and natural orbitals obtained from the one-body density matrix. This is remarkable because, generally, it is not possible to recover the wave function of a quantum many-body system solely from the one-body density matrix. The proposed ground state can be used to describe the ground state of the fermionic Tonks-Girardeau gas under any external potential, due to the universality of the coefficients $p_N$, which is a consequence of the universality of the occupation numbers $\lambda_k^{(N)}$ \cite{koscik}. By expressing the newly formulated wave function in the second quantization, as given by Eqs.~(\ref{wf N even 2n quan}) and~(\ref{wf N odd 2n quan}), we demonstrate that the proposed wave function is a specific case of a number-conserving BCS state and provide the explicit form of the $\alpha$ coefficients in Eqs.~(\ref{alpha even}) and~(\ref{alpha_odd}). Finally, in Eq.~(\ref{O3 FTG}), we explicitly derive the expectation values of the operators to be measured in order to experimentally detect pairing in the fermionic Tonks-Girardeau gas.

We expect our results to have an impact on understanding one-dimensional quantum systems, in particular the fermionic Tonks-Girardeau gas, and to be useful for the study of quantum phenomena and quantum correlations occurring in $p$-wave fermions. 
The system under study is particularly appealing as it is not a Luttinger Liquid, which instead describes compressible one-dimensional gases. Therefore, it is of great interest to understand how quantum correlations form beyond the Luttinger Liquid paradigm.
All previous studies on the fermionic Tonks-Girardeau gas were based on pair-product form~(\ref{gs attractive fermions}) \cite{GirardeauNguyenOlshanii2004,GirardeauMinguzzi2006,MinguzziGirardeau2006,BenderErkerGranger2005,Kościk_2020,koscik,delcampo2006,Hao2007}. Since our proposed wave function is well-suited for working within a second quantization framework, we anticipate that our new formulation of the ground state will lead to novel investigations on the fermionic Tonks-Girardeau gas that were not possible with the previous wave function. 
For instance, we anticipate that the proposed wave function can be used to predict and study whether pairing is also present in the excitations of the system. On the other hand, the proposed wave function is well suited for calculating relevant quantities in the study of entanglement between fermionic orbitals, as has already been done with other systems \cite{entanglement_orbitals,entanglement_orbitals2,P_rez_Obiol_2023}. Furthermore, we expect this Letter to be of interest to experimentalists, as we present the values of a set of observables for the fermionic Tonks-Girardeau gas that can be measured in order to detect pairing in fermionic systems. 

Beyond the specific relevance of this Letter for the study of the fermionic Tonks-Girardeau gas, we anticipate our findings to have a broader impact. Our results present a novel example of a system that can be exactly described using the same quantum states that describe a BCS superconductor~\cite{BCS}, which have raised a lot of interest from a wide range of fields in physics. Thus, we expect our results to be of significance to the study of BCS superconductivity. 
\begin{acknowledgments}
We acknowledge helpful and insightful discussions with Joan Martorell, Maciej Lewenstein, Utso Bhattacharya,  
Rohit Kishan Ray and Felipe Isaule that significantly contributed to the development of this Letter.

This work has been funded by Grants No.~PID2020-114626GB-I00 and PID2020-113565GB-C21 by MCIN/AEI/10.13039/5011 00011033 and 
"Unit of Excellence Mar\'ia de Maeztu 2020-2023” 
award to the Institute of Cosmos Sciences, Grant CEX2019-000918-M funded by MCIN/AEI/10.13039/501100011033. 
We acknowledge financial support from the Generalitat de Catalunya (Grants 2021SGR01411 and 2021SGR01095). 
A.R.-F. acknowledges funding from MIU through Grant No. FPU20/06174. F.S. acknowledges funding from UB through Grant Master+UB 2023.2.FFIS.1. We acknowledge funding from Grants 338
PID2023-147475NB-I00 and CEX2024-001451-M financed 339
by MCIN/AEI/10.13039/501100011033

\end{acknowledgments}

\bibliography{biblio}

\begin{thebibliography}{37}%
\makeatletter
\providecommand \@ifxundefined [1]{%
 \@ifx{#1\undefined}
}%
\providecommand \@ifnum [1]{%
 \ifnum #1\expandafter \@firstoftwo
 \else \expandafter \@secondoftwo
 \fi
}%
\providecommand \@ifx [1]{%
 \ifx #1\expandafter \@firstoftwo
 \else \expandafter \@secondoftwo
 \fi
}%
\providecommand \natexlab [1]{#1}%
\providecommand \enquote  [1]{``#1''}%
\providecommand \bibnamefont  [1]{#1}%
\providecommand \bibfnamefont [1]{#1}%
\providecommand \citenamefont [1]{#1}%
\providecommand \href@noop [0]{\@secondoftwo}%
\providecommand \href [0]{\begingroup \@sanitize@url \@href}%
\providecommand \@href[1]{\@@startlink{#1}\@@href}%
\providecommand \@@href[1]{\endgroup#1\@@endlink}%
\providecommand \@sanitize@url [0]{\catcode `\\12\catcode `\$12\catcode `\&12\catcode `\#12\catcode `\^12\catcode `\_12\catcode `\%12\relax}%
\providecommand \@@startlink[1]{}%
\providecommand \@@endlink[0]{}%
\providecommand \url  [0]{\begingroup\@sanitize@url \@url }%
\providecommand \@url [1]{\endgroup\@href {#1}{\urlprefix }}%
\providecommand \urlprefix  [0]{URL }%
\providecommand \Eprint [0]{\href }%
\providecommand \doibase [0]{https://doi.org/}%
\providecommand \selectlanguage [0]{\@gobble}%
\providecommand \bibinfo  [0]{\@secondoftwo}%
\providecommand \bibfield  [0]{\@secondoftwo}%
\providecommand \translation [1]{[#1]}%
\providecommand \BibitemOpen [0]{}%
\providecommand \bibitemStop [0]{}%
\providecommand \bibitemNoStop [0]{.\EOS\space}%
\providecommand \EOS [0]{\spacefactor3000\relax}%
\providecommand \BibitemShut  [1]{\csname bibitem#1\endcsname}%
\let\auto@bib@innerbib\@empty
\bibitem [{\citenamefont {Sutherland}(2004)}]{sutherland2004beautiful}%
  \BibitemOpen
  \bibfield  {author} {\bibinfo {author} {\bibfnamefont {B.}~\bibnamefont {Sutherland}},\ }\href {https://doi.org/10.1142/5552} {\emph {\bibinfo {title} {Beautiful Models}}}\ (\bibinfo  {publisher} {WORLD SCIENTIFIC},\ \bibinfo {year} {2004})\BibitemShut {NoStop}%
\bibitem [{\citenamefont {Lewenstein}\ \emph {et~al.}(2007)\citenamefont {Lewenstein}, \citenamefont {Sanpera}, \citenamefont {Ahufinger}, \citenamefont {Damski}, \citenamefont {Sen(De)},\ and\ \citenamefont {Sen}}]{Lewenstein_2007}%
  \BibitemOpen
  \bibfield  {author} {\bibinfo {author} {\bibfnamefont {M.}~\bibnamefont {Lewenstein}}, \bibinfo {author} {\bibfnamefont {A.}~\bibnamefont {Sanpera}}, \bibinfo {author} {\bibfnamefont {V.}~\bibnamefont {Ahufinger}}, \bibinfo {author} {\bibfnamefont {B.}~\bibnamefont {Damski}}, \bibinfo {author} {\bibfnamefont {A.}~\bibnamefont {Sen(De)}},\ and\ \bibinfo {author} {\bibfnamefont {U.}~\bibnamefont {Sen}},\ }\bibfield  {title} {\bibinfo {title} {Ultracold atomic gases in optical lattices: mimicking condensed matter physics and beyond},\ }\href {https://doi.org/10.1080/00018730701223200} {\bibfield  {journal} {\bibinfo  {journal} {Advances in Physics}\ }\textbf {\bibinfo {volume} {56}},\ \bibinfo {pages} {243–379} (\bibinfo {year} {2007})}\BibitemShut {NoStop}%
\bibitem [{\citenamefont {Greiner}\ \emph {et~al.}(2001)\citenamefont {Greiner}, \citenamefont {Bloch}, \citenamefont {Mandel}, \citenamefont {H\"ansch},\ and\ \citenamefont {Esslinger}}]{Greiner2001}%
  \BibitemOpen
  \bibfield  {author} {\bibinfo {author} {\bibfnamefont {M.}~\bibnamefont {Greiner}}, \bibinfo {author} {\bibfnamefont {I.}~\bibnamefont {Bloch}}, \bibinfo {author} {\bibfnamefont {O.}~\bibnamefont {Mandel}}, \bibinfo {author} {\bibfnamefont {T.~W.}\ \bibnamefont {H\"ansch}},\ and\ \bibinfo {author} {\bibfnamefont {T.}~\bibnamefont {Esslinger}},\ }\bibfield  {title} {\bibinfo {title} {Exploring phase coherence in a 2d lattice of bose-einstein condensates},\ }\href {https://doi.org/10.1103/PhysRevLett.87.160405} {\bibfield  {journal} {\bibinfo  {journal} {Phys. Rev. Lett.}\ }\textbf {\bibinfo {volume} {87}},\ \bibinfo {pages} {160405} (\bibinfo {year} {2001})}\BibitemShut {NoStop}%
\bibitem [{\citenamefont {G\"orlitz}\ \emph {et~al.}(2001)\citenamefont {G\"orlitz}, \citenamefont {Vogels}, \citenamefont {Leanhardt}, \citenamefont {Raman}, \citenamefont {Gustavson}, \citenamefont {Abo-Shaeer}, \citenamefont {Chikkatur}, \citenamefont {Gupta}, \citenamefont {Inouye}, \citenamefont {Rosenband},\ and\ \citenamefont {Ketterle}}]{Gorlitz2001}%
  \BibitemOpen
  \bibfield  {author} {\bibinfo {author} {\bibfnamefont {A.}~\bibnamefont {G\"orlitz}}, \bibinfo {author} {\bibfnamefont {J.~M.}\ \bibnamefont {Vogels}}, \bibinfo {author} {\bibfnamefont {A.~E.}\ \bibnamefont {Leanhardt}}, \bibinfo {author} {\bibfnamefont {C.}~\bibnamefont {Raman}}, \bibinfo {author} {\bibfnamefont {T.~L.}\ \bibnamefont {Gustavson}}, \bibinfo {author} {\bibfnamefont {J.~R.}\ \bibnamefont {Abo-Shaeer}}, \bibinfo {author} {\bibfnamefont {A.~P.}\ \bibnamefont {Chikkatur}}, \bibinfo {author} {\bibfnamefont {S.}~\bibnamefont {Gupta}}, \bibinfo {author} {\bibfnamefont {S.}~\bibnamefont {Inouye}}, \bibinfo {author} {\bibfnamefont {T.}~\bibnamefont {Rosenband}},\ and\ \bibinfo {author} {\bibfnamefont {W.}~\bibnamefont {Ketterle}},\ }\bibfield  {title} {\bibinfo {title} {Realization of bose-einstein condensates in lower dimensions},\ }\href {https://doi.org/10.1103/PhysRevLett.87.130402} {\bibfield  {journal} {\bibinfo  {journal} {Phys. Rev. Lett.}\ }\textbf {\bibinfo {volume} {87}},\ \bibinfo {pages}
  {130402} (\bibinfo {year} {2001})}\BibitemShut {NoStop}%
\bibitem [{\citenamefont {Haller}\ \emph {et~al.}(2009)\citenamefont {Haller}, \citenamefont {Gustavsson}, \citenamefont {Mark}, \citenamefont {Danzl}, \citenamefont {Hart}, \citenamefont {Pupillo},\ and\ \citenamefont {Nägerl}}]{HallerNägerl09}%
  \BibitemOpen
  \bibfield  {author} {\bibinfo {author} {\bibfnamefont {E.}~\bibnamefont {Haller}}, \bibinfo {author} {\bibfnamefont {M.}~\bibnamefont {Gustavsson}}, \bibinfo {author} {\bibfnamefont {M.~J.}\ \bibnamefont {Mark}}, \bibinfo {author} {\bibfnamefont {J.~G.}\ \bibnamefont {Danzl}}, \bibinfo {author} {\bibfnamefont {R.}~\bibnamefont {Hart}}, \bibinfo {author} {\bibfnamefont {G.}~\bibnamefont {Pupillo}},\ and\ \bibinfo {author} {\bibfnamefont {H.-C.}\ \bibnamefont {Nägerl}},\ }\bibfield  {title} {\bibinfo {title} {Realization of an excited, strongly correlated quantum gas phase},\ }\href {https://doi.org/10.1126/science.1175850} {\bibfield  {journal} {\bibinfo  {journal} {Science}\ }\textbf {\bibinfo {volume} {325}},\ \bibinfo {pages} {1224} (\bibinfo {year} {2009})}\BibitemShut {NoStop}%
\bibitem [{\citenamefont {Kao}\ \emph {et~al.}(2021)\citenamefont {Kao}, \citenamefont {Li}, \citenamefont {Lin}, \citenamefont {Gopalakrishnan},\ and\ \citenamefont {Lev}}]{KaoLev21}%
  \BibitemOpen
  \bibfield  {author} {\bibinfo {author} {\bibfnamefont {W.}~\bibnamefont {Kao}}, \bibinfo {author} {\bibfnamefont {K.-Y.}\ \bibnamefont {Li}}, \bibinfo {author} {\bibfnamefont {K.-Y.}\ \bibnamefont {Lin}}, \bibinfo {author} {\bibfnamefont {S.}~\bibnamefont {Gopalakrishnan}},\ and\ \bibinfo {author} {\bibfnamefont {B.~L.}\ \bibnamefont {Lev}},\ }\bibfield  {title} {\bibinfo {title} {Topological pumping of a 1d dipolar gas into strongly correlated prethermal states},\ }\href {https://doi.org/10.1126/science.abb4928} {\bibfield  {journal} {\bibinfo  {journal} {Science}\ }\textbf {\bibinfo {volume} {371}},\ \bibinfo {pages} {296} (\bibinfo {year} {2021})}\BibitemShut {NoStop}%
\bibitem [{\citenamefont {Z\"urn}\ \emph {et~al.}(2012)\citenamefont {Z\"urn}, \citenamefont {Serwane}, \citenamefont {Lompe}, \citenamefont {Wenz}, \citenamefont {Ries}, \citenamefont {Bohn},\ and\ \citenamefont {Jochim}}]{Zurn2012}%
  \BibitemOpen
  \bibfield  {author} {\bibinfo {author} {\bibfnamefont {G.}~\bibnamefont {Z\"urn}}, \bibinfo {author} {\bibfnamefont {F.}~\bibnamefont {Serwane}}, \bibinfo {author} {\bibfnamefont {T.}~\bibnamefont {Lompe}}, \bibinfo {author} {\bibfnamefont {A.~N.}\ \bibnamefont {Wenz}}, \bibinfo {author} {\bibfnamefont {M.~G.}\ \bibnamefont {Ries}}, \bibinfo {author} {\bibfnamefont {J.~E.}\ \bibnamefont {Bohn}},\ and\ \bibinfo {author} {\bibfnamefont {S.}~\bibnamefont {Jochim}},\ }\bibfield  {title} {\bibinfo {title} {Fermionization of two distinguishable fermions},\ }\href {https://doi.org/10.1103/PhysRevLett.108.075303} {\bibfield  {journal} {\bibinfo  {journal} {Phys. Rev. Lett.}\ }\textbf {\bibinfo {volume} {108}},\ \bibinfo {pages} {075303} (\bibinfo {year} {2012})}\BibitemShut {NoStop}%
\bibitem [{\citenamefont {Wenz}\ \emph {et~al.}(2013)\citenamefont {Wenz}, \citenamefont {Zürn}, \citenamefont {Murmann}, \citenamefont {Brouzos}, \citenamefont {Lompe},\ and\ \citenamefont {Jochim}}]{Wenz_2013}%
  \BibitemOpen
  \bibfield  {author} {\bibinfo {author} {\bibfnamefont {A.~N.}\ \bibnamefont {Wenz}}, \bibinfo {author} {\bibfnamefont {G.}~\bibnamefont {Zürn}}, \bibinfo {author} {\bibfnamefont {S.}~\bibnamefont {Murmann}}, \bibinfo {author} {\bibfnamefont {I.}~\bibnamefont {Brouzos}}, \bibinfo {author} {\bibfnamefont {T.}~\bibnamefont {Lompe}},\ and\ \bibinfo {author} {\bibfnamefont {S.}~\bibnamefont {Jochim}},\ }\bibfield  {title} {\bibinfo {title} {From few to many: Observing the formation of a fermi sea one atom at a time},\ }\href {https://doi.org/10.1126/science.1240516} {\bibfield  {journal} {\bibinfo  {journal} {Science}\ }\textbf {\bibinfo {volume} {342}},\ \bibinfo {pages} {457–460} (\bibinfo {year} {2013})}\BibitemShut {NoStop}%
\bibitem [{\citenamefont {Granger}\ and\ \citenamefont {Blume}(2004)}]{GrangerBlume}%
  \BibitemOpen
  \bibfield  {author} {\bibinfo {author} {\bibfnamefont {B.~E.}\ \bibnamefont {Granger}}\ and\ \bibinfo {author} {\bibfnamefont {D.}~\bibnamefont {Blume}},\ }\bibfield  {title} {\bibinfo {title} {Tuning the interactions of spin-polarized fermions using quasi-one-dimensional confinement},\ }\href {https://doi.org/10.1103/PhysRevLett.92.133202} {\bibfield  {journal} {\bibinfo  {journal} {Phys. Rev. Lett.}\ }\textbf {\bibinfo {volume} {92}},\ \bibinfo {pages} {133202} (\bibinfo {year} {2004})}\BibitemShut {NoStop}%
\bibitem [{\citenamefont {Girardeau}(1960)}]{girardeau_mapping}%
  \BibitemOpen
  \bibfield  {author} {\bibinfo {author} {\bibfnamefont {M.}~\bibnamefont {Girardeau}},\ }\bibfield  {title} {\bibinfo {title} {{Relationship between Systems of Impenetrable Bosons and Fermions in One Dimension}},\ }\href {https://doi.org/10.1063/1.1703687} {\bibfield  {journal} {\bibinfo  {journal} {Journal of Mathematical Physics}\ }\textbf {\bibinfo {volume} {1}},\ \bibinfo {pages} {516} (\bibinfo {year} {1960})}\BibitemShut {NoStop}%
\bibitem [{\citenamefont {Cheon}\ and\ \citenamefont {Shigehara}(1999)}]{Cheon1999}%
  \BibitemOpen
  \bibfield  {author} {\bibinfo {author} {\bibfnamefont {T.}~\bibnamefont {Cheon}}\ and\ \bibinfo {author} {\bibfnamefont {T.}~\bibnamefont {Shigehara}},\ }\bibfield  {title} {\bibinfo {title} {Fermion-boson duality of one-dimensional quantum particles with generalized contact interactions},\ }\href {https://doi.org/10.1103/PhysRevLett.82.2536} {\bibfield  {journal} {\bibinfo  {journal} {Phys. Rev. Lett.}\ }\textbf {\bibinfo {volume} {82}},\ \bibinfo {pages} {2536} (\bibinfo {year} {1999})}\BibitemShut {NoStop}%
\bibitem [{\citenamefont {Sekino}\ \emph {et~al.}(2018)\citenamefont {Sekino}, \citenamefont {Tan},\ and\ \citenamefont {Nishida}}]{Sekino}%
  \BibitemOpen
  \bibfield  {author} {\bibinfo {author} {\bibfnamefont {Y.}~\bibnamefont {Sekino}}, \bibinfo {author} {\bibfnamefont {S.}~\bibnamefont {Tan}},\ and\ \bibinfo {author} {\bibfnamefont {Y.}~\bibnamefont {Nishida}},\ }\bibfield  {title} {\bibinfo {title} {Comparative study of one-dimensional bose and fermi gases with contact interactions from the viewpoint of universal relations for correlation functions},\ }\href {https://doi.org/10.1103/PhysRevA.97.013621} {\bibfield  {journal} {\bibinfo  {journal} {Phys. Rev. A}\ }\textbf {\bibinfo {volume} {97}},\ \bibinfo {pages} {013621} (\bibinfo {year} {2018})}\BibitemShut {NoStop}%
\bibitem [{\citenamefont {Girardeau}\ \emph {et~al.}(2001)\citenamefont {Girardeau}, \citenamefont {Wright},\ and\ \citenamefont {Triscari}}]{Girardeau2001}%
  \BibitemOpen
  \bibfield  {author} {\bibinfo {author} {\bibfnamefont {M.~D.}\ \bibnamefont {Girardeau}}, \bibinfo {author} {\bibfnamefont {E.~M.}\ \bibnamefont {Wright}},\ and\ \bibinfo {author} {\bibfnamefont {J.~M.}\ \bibnamefont {Triscari}},\ }\bibfield  {title} {\bibinfo {title} {Ground-state properties of a one-dimensional system of hard-core bosons in a harmonic trap},\ }\href {https://doi.org/10.1103/PhysRevA.63.033601} {\bibfield  {journal} {\bibinfo  {journal} {Phys. Rev. A}\ }\textbf {\bibinfo {volume} {63}},\ \bibinfo {pages} {033601} (\bibinfo {year} {2001})}\BibitemShut {NoStop}%
\bibitem [{\citenamefont {Lieb}\ and\ \citenamefont {Liniger}(1963)}]{linieger}%
  \BibitemOpen
  \bibfield  {author} {\bibinfo {author} {\bibfnamefont {E.~H.}\ \bibnamefont {Lieb}}\ and\ \bibinfo {author} {\bibfnamefont {W.}~\bibnamefont {Liniger}},\ }\bibfield  {title} {\bibinfo {title} {Exact analysis of an interacting bose gas. i. the general solution and the ground state},\ }\href {https://doi.org/10.1103/PhysRev.130.1605} {\bibfield  {journal} {\bibinfo  {journal} {Phys. Rev.}\ }\textbf {\bibinfo {volume} {130}},\ \bibinfo {pages} {1605} (\bibinfo {year} {1963})}\BibitemShut {NoStop}%
\bibitem [{\citenamefont {Paredes}\ \emph {et~al.}(2004)\citenamefont {Paredes}, \citenamefont {Widera}, \citenamefont {Murg}, \citenamefont {Mandel}, \citenamefont {F{\"o}lling}, \citenamefont {Cirac}, \citenamefont {Shlyapnikov}, \citenamefont {H{\"a}nsch},\ and\ \citenamefont {Bloch}}]{paredes2004tonks}%
  \BibitemOpen
  \bibfield  {author} {\bibinfo {author} {\bibfnamefont {B.}~\bibnamefont {Paredes}}, \bibinfo {author} {\bibfnamefont {A.}~\bibnamefont {Widera}}, \bibinfo {author} {\bibfnamefont {V.}~\bibnamefont {Murg}}, \bibinfo {author} {\bibfnamefont {O.}~\bibnamefont {Mandel}}, \bibinfo {author} {\bibfnamefont {S.}~\bibnamefont {F{\"o}lling}}, \bibinfo {author} {\bibfnamefont {I.}~\bibnamefont {Cirac}}, \bibinfo {author} {\bibfnamefont {G.~V.}\ \bibnamefont {Shlyapnikov}}, \bibinfo {author} {\bibfnamefont {T.~W.}\ \bibnamefont {H{\"a}nsch}},\ and\ \bibinfo {author} {\bibfnamefont {I.}~\bibnamefont {Bloch}},\ }\bibfield  {title} {\bibinfo {title} {Tonks–girardeau gas of ultracold atoms in an optical lattice},\ }\href {https://doi.org/10.1038/nature02530} {\bibfield  {journal} {\bibinfo  {journal} {Nature}\ }\textbf {\bibinfo {volume} {429}},\ \bibinfo {pages} {277} (\bibinfo {year} {2004})}\BibitemShut {NoStop}%
\bibitem [{\citenamefont {Marmorini}\ \emph {et~al.}(2016)\citenamefont {Marmorini}, \citenamefont {Pepe},\ and\ \citenamefont {Calabrese}}]{Marmorini_2016}%
  \BibitemOpen
  \bibfield  {author} {\bibinfo {author} {\bibfnamefont {G.}~\bibnamefont {Marmorini}}, \bibinfo {author} {\bibfnamefont {M.}~\bibnamefont {Pepe}},\ and\ \bibinfo {author} {\bibfnamefont {P.}~\bibnamefont {Calabrese}},\ }\bibfield  {title} {\bibinfo {title} {One-body reduced density matrix of trapped impenetrable anyons in one dimension},\ }\href {https://doi.org/10.1088/1742-5468/2016/07/073106} {\bibfield  {journal} {\bibinfo  {journal} {Journal of Statistical Mechanics: Theory and Experiment}\ }\textbf {\bibinfo {volume} {2016}},\ \bibinfo {pages} {073106} (\bibinfo {year} {2016})}\BibitemShut {NoStop}%
\bibitem [{\citenamefont {Yukalov}\ and\ \citenamefont {Girardeau}(2005)}]{Yukalov_2005}%
  \BibitemOpen
  \bibfield  {author} {\bibinfo {author} {\bibfnamefont {V.~I.}\ \bibnamefont {Yukalov}}\ and\ \bibinfo {author} {\bibfnamefont {M.~D.}\ \bibnamefont {Girardeau}},\ }\bibfield  {title} {\bibinfo {title} {Fermi-bose mapping for one-dimensional bose gases},\ }\href {https://doi.org/10.1002/lapl.200510011} {\bibfield  {journal} {\bibinfo  {journal} {Laser Physics Letters}\ }\textbf {\bibinfo {volume} {2}},\ \bibinfo {pages} {375} (\bibinfo {year} {2005})}\BibitemShut {NoStop}%
\bibitem [{\citenamefont {Girardeau}\ \emph {et~al.}(2004)\citenamefont {Girardeau}, \citenamefont {Nguyen},\ and\ \citenamefont {Olshanii}}]{GirardeauNguyenOlshanii2004}%
  \BibitemOpen
  \bibfield  {author} {\bibinfo {author} {\bibfnamefont {M.~D.}\ \bibnamefont {Girardeau}}, \bibinfo {author} {\bibfnamefont {H.}~\bibnamefont {Nguyen}},\ and\ \bibinfo {author} {\bibfnamefont {M.}~\bibnamefont {Olshanii}},\ }\bibfield  {title} {\bibinfo {title} {Effective interactions, fermi–bose duality, and ground states of ultracold atomic vapors in tight de broglie waveguides},\ }\href {https://doi.org/10.1016/j.optcom.2004.09.079} {\bibfield  {journal} {\bibinfo  {journal} {Opt. Commun.}\ }\textbf {\bibinfo {volume} {243}},\ \bibinfo {pages} {3} (\bibinfo {year} {2004})}\BibitemShut {NoStop}%
\bibitem [{\citenamefont {Girardeau}\ and\ \citenamefont {Minguzzi}(2006)}]{GirardeauMinguzzi2006}%
  \BibitemOpen
  \bibfield  {author} {\bibinfo {author} {\bibfnamefont {M.~D.}\ \bibnamefont {Girardeau}}\ and\ \bibinfo {author} {\bibfnamefont {A.}~\bibnamefont {Minguzzi}},\ }\bibfield  {title} {\bibinfo {title} {Bosonization, pairing, and superconductivity of the fermionic tonks-girardeau gas},\ }\href {https://doi.org/10.1103/PhysRevLett.96.080404} {\bibfield  {journal} {\bibinfo  {journal} {Phys. Rev. Lett.}\ }\textbf {\bibinfo {volume} {96}},\ \bibinfo {pages} {080404} (\bibinfo {year} {2006})}\BibitemShut {NoStop}%
\bibitem [{\citenamefont {Minguzzi}\ and\ \citenamefont {Girardeau}(2006)}]{MinguzziGirardeau2006}%
  \BibitemOpen
  \bibfield  {author} {\bibinfo {author} {\bibfnamefont {A.}~\bibnamefont {Minguzzi}}\ and\ \bibinfo {author} {\bibfnamefont {M.~D.}\ \bibnamefont {Girardeau}},\ }\bibfield  {title} {\bibinfo {title} {Pairing of a harmonically trapped fermionic tonks-girardeau gas},\ }\href {https://doi.org/10.1103/PhysRevA.73.063614} {\bibfield  {journal} {\bibinfo  {journal} {Phys. Rev. A}\ }\textbf {\bibinfo {volume} {73}},\ \bibinfo {pages} {063614} (\bibinfo {year} {2006})}\BibitemShut {NoStop}%
\bibitem [{\citenamefont {Bender}\ \emph {et~al.}(2005)\citenamefont {Bender}, \citenamefont {Erker},\ and\ \citenamefont {Granger}}]{BenderErkerGranger2005}%
  \BibitemOpen
  \bibfield  {author} {\bibinfo {author} {\bibfnamefont {S.~A.}\ \bibnamefont {Bender}}, \bibinfo {author} {\bibfnamefont {K.~D.}\ \bibnamefont {Erker}},\ and\ \bibinfo {author} {\bibfnamefont {B.~E.}\ \bibnamefont {Granger}},\ }\bibfield  {title} {\bibinfo {title} {Exponentially decaying correlations in a gas of strongly interacting spin-polarized 1d fermions with zero-range interactions},\ }\href {https://doi.org/10.1103/PhysRevLett.95.230404} {\bibfield  {journal} {\bibinfo  {journal} {Phys. Rev. Lett.}\ }\textbf {\bibinfo {volume} {95}},\ \bibinfo {pages} {230404} (\bibinfo {year} {2005})}\BibitemShut {NoStop}%
\bibitem [{\citenamefont {Kościk}\ and\ \citenamefont {Sowiński}(2020)}]{Kościk_2020}%
  \BibitemOpen
  \bibfield  {author} {\bibinfo {author} {\bibfnamefont {P.}~\bibnamefont {Kościk}}\ and\ \bibinfo {author} {\bibfnamefont {T.}~\bibnamefont {Sowiński}},\ }\bibfield  {title} {\bibinfo {title} {Variational ansatz for p-wave fermions confined in a one-dimensional harmonic trap},\ }\href {https://doi.org/10.1088/1367-2630/abb386} {\bibfield  {journal} {\bibinfo  {journal} {New Journal of Physics}\ }\textbf {\bibinfo {volume} {22}},\ \bibinfo {pages} {093053} (\bibinfo {year} {2020})}\BibitemShut {NoStop}%
\bibitem [{\citenamefont {Ko\ifmmode~\acute{s}\else \'{s}\fi{}cik}\ and\ \citenamefont {Sowi\ifmmode~\acute{n}\else \'{n}\fi{}ski}(2023)}]{koscik}%
  \BibitemOpen
  \bibfield  {author} {\bibinfo {author} {\bibfnamefont {P.}~\bibnamefont {Ko\ifmmode~\acute{s}\else \'{s}\fi{}cik}}\ and\ \bibinfo {author} {\bibfnamefont {T.}~\bibnamefont {Sowi\ifmmode~\acute{n}\else \'{n}\fi{}ski}},\ }\bibfield  {title} {\bibinfo {title} {Universality of internal correlations of strongly interacting $p$-wave fermions in one-dimensional geometry},\ }\href {https://doi.org/10.1103/PhysRevLett.130.253401} {\bibfield  {journal} {\bibinfo  {journal} {Phys. Rev. Lett.}\ }\textbf {\bibinfo {volume} {130}},\ \bibinfo {pages} {253401} (\bibinfo {year} {2023})}\BibitemShut {NoStop}%
\bibitem [{\citenamefont {del Campo}\ \emph {et~al.}(2006)\citenamefont {del Campo}, \citenamefont {Delgado}, \citenamefont {Garc\'{\i}a-Calder\'on}, \citenamefont {Muga},\ and\ \citenamefont {Raizen}}]{delcampo2006}%
  \BibitemOpen
  \bibfield  {author} {\bibinfo {author} {\bibfnamefont {A.}~\bibnamefont {del Campo}}, \bibinfo {author} {\bibfnamefont {F.}~\bibnamefont {Delgado}}, \bibinfo {author} {\bibfnamefont {G.}~\bibnamefont {Garc\'{\i}a-Calder\'on}}, \bibinfo {author} {\bibfnamefont {J.~G.}\ \bibnamefont {Muga}},\ and\ \bibinfo {author} {\bibfnamefont {M.~G.}\ \bibnamefont {Raizen}},\ }\bibfield  {title} {\bibinfo {title} {Decay by tunneling of bosonic and fermionic tonks-girardeau gases},\ }\href {https://doi.org/10.1103/PhysRevA.74.013605} {\bibfield  {journal} {\bibinfo  {journal} {Phys. Rev. A}\ }\textbf {\bibinfo {volume} {74}},\ \bibinfo {pages} {013605} (\bibinfo {year} {2006})}\BibitemShut {NoStop}%
\bibitem [{\citenamefont {Hao}\ \emph {et~al.}(2007)\citenamefont {Hao}, \citenamefont {Zhang},\ and\ \citenamefont {Chen}}]{Hao2007}%
  \BibitemOpen
  \bibfield  {author} {\bibinfo {author} {\bibfnamefont {Y.}~\bibnamefont {Hao}}, \bibinfo {author} {\bibfnamefont {Y.}~\bibnamefont {Zhang}},\ and\ \bibinfo {author} {\bibfnamefont {S.}~\bibnamefont {Chen}},\ }\bibfield  {title} {\bibinfo {title} {One-dimensional fermionic gases with attractive $p$-wave interaction in a hard-wall trap},\ }\href {https://doi.org/10.1103/PhysRevA.76.063601} {\bibfield  {journal} {\bibinfo  {journal} {Phys. Rev. A}\ }\textbf {\bibinfo {volume} {76}},\ \bibinfo {pages} {063601} (\bibinfo {year} {2007})}\BibitemShut {NoStop}%
\bibitem [{\citenamefont {Sabater}\ \emph {et~al.}(2024)\citenamefont {Sabater}, \citenamefont {Rojo-Franc\`as}, \citenamefont {Astrakharchik},\ and\ \citenamefont {Juli\'a-D\'{\i}az}}]{Sabater}%
  \BibitemOpen
  \bibfield  {author} {\bibinfo {author} {\bibfnamefont {F.}~\bibnamefont {Sabater}}, \bibinfo {author} {\bibfnamefont {A.}~\bibnamefont {Rojo-Franc\`as}}, \bibinfo {author} {\bibfnamefont {G.~E.}\ \bibnamefont {Astrakharchik}},\ and\ \bibinfo {author} {\bibfnamefont {B.}~\bibnamefont {Juli\'a-D\'{\i}az}},\ }\bibfield  {title} {\bibinfo {title} {Universal composite boson formation in strongly interacting one-dimensional fermionic systems},\ }\href {https://doi.org/10.1103/PhysRevLett.132.193401} {\bibfield  {journal} {\bibinfo  {journal} {Phys. Rev. Lett.}\ }\textbf {\bibinfo {volume} {132}},\ \bibinfo {pages} {193401} (\bibinfo {year} {2024})}\BibitemShut {NoStop}%
\bibitem [{\citenamefont {Bardeen}\ \emph {et~al.}(1957)\citenamefont {Bardeen}, \citenamefont {Cooper},\ and\ \citenamefont {Schrieffer}}]{BCS}%
  \BibitemOpen
  \bibfield  {author} {\bibinfo {author} {\bibfnamefont {J.}~\bibnamefont {Bardeen}}, \bibinfo {author} {\bibfnamefont {L.~N.}\ \bibnamefont {Cooper}},\ and\ \bibinfo {author} {\bibfnamefont {J.~R.}\ \bibnamefont {Schrieffer}},\ }\bibfield  {title} {\bibinfo {title} {Theory of superconductivity},\ }\href {https://doi.org/10.1103/PhysRev.108.1175} {\bibfield  {journal} {\bibinfo  {journal} {Phys. Rev.}\ }\textbf {\bibinfo {volume} {108}},\ \bibinfo {pages} {1175} (\bibinfo {year} {1957})}\BibitemShut {NoStop}%
\bibitem [{\citenamefont {Kraus}\ \emph {et~al.}(2009)\citenamefont {Kraus}, \citenamefont {Wolf}, \citenamefont {Cirac},\ and\ \citenamefont {Giedke}}]{pairing_krauss2009}%
  \BibitemOpen
  \bibfield  {author} {\bibinfo {author} {\bibfnamefont {C.~V.}\ \bibnamefont {Kraus}}, \bibinfo {author} {\bibfnamefont {M.~M.}\ \bibnamefont {Wolf}}, \bibinfo {author} {\bibfnamefont {J.~I.}\ \bibnamefont {Cirac}},\ and\ \bibinfo {author} {\bibfnamefont {G.}~\bibnamefont {Giedke}},\ }\bibfield  {title} {\bibinfo {title} {Pairing in fermionic systems: A quantum-information perspective},\ }\href {https://doi.org/10.1103/PhysRevA.79.012306} {\bibfield  {journal} {\bibinfo  {journal} {Phys. Rev. A}\ }\textbf {\bibinfo {volume} {79}},\ \bibinfo {pages} {012306} (\bibinfo {year} {2009})}\BibitemShut {NoStop}%
\bibitem [{sup()}]{supplemental}%
  \BibitemOpen
  \href@noop {} {}\bibinfo {note} {See Supplemental Material for the proofs of $\psi_T=\psi_F$; and for detailed calculation of $\langle\vec{O}_3^{FTG}\rangle$}\BibitemShut {NoStop}%
\bibitem [{\citenamefont {Schliemann}\ \emph {et~al.}(2001)\citenamefont {Schliemann}, \citenamefont {Cirac}, \citenamefont {Ku\ifmmode~\acute{s}\else \'{s}\fi{}}, \citenamefont {Lewenstein},\ and\ \citenamefont {Loss}}]{correlation2fermions}%
  \BibitemOpen
  \bibfield  {author} {\bibinfo {author} {\bibfnamefont {J.}~\bibnamefont {Schliemann}}, \bibinfo {author} {\bibfnamefont {J.~I.}\ \bibnamefont {Cirac}}, \bibinfo {author} {\bibfnamefont {M.}~\bibnamefont {Ku\ifmmode~\acute{s}\else \'{s}\fi{}}}, \bibinfo {author} {\bibfnamefont {M.}~\bibnamefont {Lewenstein}},\ and\ \bibinfo {author} {\bibfnamefont {D.}~\bibnamefont {Loss}},\ }\bibfield  {title} {\bibinfo {title} {Quantum correlations in two-fermion systems},\ }\href {https://doi.org/10.1103/PhysRevA.64.022303} {\bibfield  {journal} {\bibinfo  {journal} {Phys. Rev. A}\ }\textbf {\bibinfo {volume} {64}},\ \bibinfo {pages} {022303} (\bibinfo {year} {2001})}\BibitemShut {NoStop}%
\bibitem [{\citenamefont {Eckert}\ \emph {et~al.}(2002)\citenamefont {Eckert}, \citenamefont {Schliemann}, \citenamefont {Bruß},\ and\ \citenamefont {Lewenstein}}]{Eckert_2002}%
  \BibitemOpen
  \bibfield  {author} {\bibinfo {author} {\bibfnamefont {K.}~\bibnamefont {Eckert}}, \bibinfo {author} {\bibfnamefont {J.}~\bibnamefont {Schliemann}}, \bibinfo {author} {\bibfnamefont {D.}~\bibnamefont {Bruß}},\ and\ \bibinfo {author} {\bibfnamefont {M.}~\bibnamefont {Lewenstein}},\ }\bibfield  {title} {\bibinfo {title} {Quantum correlations in systems of indistinguishable particles},\ }\href {https://doi.org/10.1006/aphy.2002.6268} {\bibfield  {journal} {\bibinfo  {journal} {Annals of Physics}\ }\textbf {\bibinfo {volume} {299}},\ \bibinfo {pages} {88–127} (\bibinfo {year} {2002})}\BibitemShut {NoStop}%
\bibitem [{\citenamefont {Sekino}\ and\ \citenamefont {Nishida}(2021)}]{three_body_1}%
  \BibitemOpen
  \bibfield  {author} {\bibinfo {author} {\bibfnamefont {Y.}~\bibnamefont {Sekino}}\ and\ \bibinfo {author} {\bibfnamefont {Y.}~\bibnamefont {Nishida}},\ }\bibfield  {title} {\bibinfo {title} {Field-theoretical aspects of one-dimensional bose and fermi gases with contact interactions},\ }\href {https://doi.org/10.1103/PhysRevA.103.043307} {\bibfield  {journal} {\bibinfo  {journal} {Phys. Rev. A}\ }\textbf {\bibinfo {volume} {103}},\ \bibinfo {pages} {043307} (\bibinfo {year} {2021})}\BibitemShut {NoStop}%
\bibitem [{\citenamefont {Valiente}(2021)}]{three_body_2}%
  \BibitemOpen
  \bibfield  {author} {\bibinfo {author} {\bibfnamefont {M.}~\bibnamefont {Valiente}},\ }\bibfield  {title} {\bibinfo {title} {Universal duality transformations in interacting one-dimensional quantum systems},\ }\href {https://doi.org/10.1103/PhysRevA.103.L021302} {\bibfield  {journal} {\bibinfo  {journal} {Phys. Rev. A}\ }\textbf {\bibinfo {volume} {103}},\ \bibinfo {pages} {L021302} (\bibinfo {year} {2021})}\BibitemShut {NoStop}%
\bibitem [{\citenamefont {Guo}\ and\ \citenamefont {Tajima}(2022)}]{three_body_3}%
  \BibitemOpen
  \bibfield  {author} {\bibinfo {author} {\bibfnamefont {Y.}~\bibnamefont {Guo}}\ and\ \bibinfo {author} {\bibfnamefont {H.}~\bibnamefont {Tajima}},\ }\bibfield  {title} {\bibinfo {title} {Stability against three-body clustering in one-dimensional spinless $p$-wave fermions},\ }\href {https://doi.org/10.1103/PhysRevA.106.043310} {\bibfield  {journal} {\bibinfo  {journal} {Phys. Rev. A}\ }\textbf {\bibinfo {volume} {106}},\ \bibinfo {pages} {043310} (\bibinfo {year} {2022})}\BibitemShut {NoStop}%
\bibitem [{\citenamefont {Shi}(2003)}]{entanglement_orbitals}%
  \BibitemOpen
  \bibfield  {author} {\bibinfo {author} {\bibfnamefont {Y.}~\bibnamefont {Shi}},\ }\bibfield  {title} {\bibinfo {title} {Quantum entanglement of identical particles},\ }\href {https://doi.org/10.1103/PhysRevA.67.024301} {\bibfield  {journal} {\bibinfo  {journal} {Phys. Rev. A}\ }\textbf {\bibinfo {volume} {67}},\ \bibinfo {pages} {024301} (\bibinfo {year} {2003})}\BibitemShut {NoStop}%
\bibitem [{\citenamefont {Gigena}\ and\ \citenamefont {Rossignoli}(2015)}]{entanglement_orbitals2}%
  \BibitemOpen
  \bibfield  {author} {\bibinfo {author} {\bibfnamefont {N.}~\bibnamefont {Gigena}}\ and\ \bibinfo {author} {\bibfnamefont {R.}~\bibnamefont {Rossignoli}},\ }\bibfield  {title} {\bibinfo {title} {Entanglement in fermion systems},\ }\href {https://doi.org/10.1103/PhysRevA.92.042326} {\bibfield  {journal} {\bibinfo  {journal} {Phys. Rev. A}\ }\textbf {\bibinfo {volume} {92}},\ \bibinfo {pages} {042326} (\bibinfo {year} {2015})}\BibitemShut {NoStop}%
\bibitem [{\citenamefont {Pérez-Obiol}\ \emph {et~al.}(2023)\citenamefont {Pérez-Obiol}, \citenamefont {Masot-Llima}, \citenamefont {Romero}, \citenamefont {Menéndez}, \citenamefont {Rios}, \citenamefont {García-Sáez},\ and\ \citenamefont {Juliá-Díaz}}]{P_rez_Obiol_2023}%
  \BibitemOpen
  \bibfield  {author} {\bibinfo {author} {\bibfnamefont {A.}~\bibnamefont {Pérez-Obiol}}, \bibinfo {author} {\bibfnamefont {S.}~\bibnamefont {Masot-Llima}}, \bibinfo {author} {\bibfnamefont {A.~M.}\ \bibnamefont {Romero}}, \bibinfo {author} {\bibfnamefont {J.}~\bibnamefont {Menéndez}}, \bibinfo {author} {\bibfnamefont {A.}~\bibnamefont {Rios}}, \bibinfo {author} {\bibfnamefont {A.}~\bibnamefont {García-Sáez}},\ and\ \bibinfo {author} {\bibfnamefont {B.}~\bibnamefont {Juliá-Díaz}},\ }\bibfield  {title} {\bibinfo {title} {Quantum entanglement patterns in the structure of atomic nuclei within the nuclear shell model},\ }\href {http://dx.doi.org/10.1140/epja/s10050-023-01151-z} {\bibfield  {journal} {\bibinfo  {journal} {The European Physical Journal A}\ }\textbf {\bibinfo {volume} {59}} (\bibinfo {year} {2023})}\BibitemShut {NoStop}%
\end{thebibliography}%
\bibliographystyle{apsrev4-2}

\appendix
\widetext{
\section{Supplemental Material}
\boldmath
\section{Proof of $\psi_T=\psi_F$ for the $N=3$ case}
\unboldmath
\noindent
In this section we prove the equality of $\psi_T=\psi_F$ for the $N=3$ case. 
In the $N=3$ case $\psi_F$ is written as 
\begin{equation}
    \psi_F(x_1,x_2,x_3)=\phi(x_1)\phi(x_2)\phi(x_3) \sign{(x_1-x_2)}\sign{(x_1-x_3)}\sign{(x_2-x_3)}. 
\end{equation}
The postulated form $\psi_T$ is
\begin{equation}
    \psi_T(x_1,x_2,x_3)=\sum_{k=1} \sqrt{\lambda_k^{(3)}} \varphi_{0,k}(x_1,x_2,x_3),
\end{equation}
where $\varphi_{0,k}(x_1,x_2,x_3)$ is the Slater determinant formed by the states $\chi_{k+}$, $\chi_{k-}$ and $\phi$. We start the demonstration by explicitly writing the form of $\varphi_{0,k}(x_1,x_2,x_3)$, 
\begin{equation}
    \frac{\psi_T(x_1, x_2, x_3)}{\phi(x_1)\phi(x_2)\phi(x_3)} =  \frac{2}{\sqrt{3!}} \sum_{k=1}\sqrt{\lambda_k^{(3)}}\bigg(\!\!\sin\big[2\pi k \big(F(x_1) - F(x_2)\big)\big] + \sin\big[2\pi k \big(F(x_2) - F(x_3)\big)\big] + \sin\big[2\pi k \big(F(x_3) - F(x_1)\big)\big]\bigg).
\end{equation}
Applying $\lambda_k^{(3)}=\frac{24}{[\pi2k]^2}$, 

\begin{equation}
    \frac{\psi_T(x_1, x_2, x_3)}{\phi(x_1)\phi(x_2)\phi(x_3)} =4\sum_{k=1}\frac{\sin\big[2\pi k \big(F(x_1) - F(x_2)\big)\big]}{2\pi k} + \frac{\sin\big[2\pi k \big(F(x_2) - F(x_3)\big)\big]}{2\pi k} + \frac{\sin\big[2\pi k \big(F(x_3) - F(x_1)\big)\big]}{2\pi k}.
\end{equation}
We now make use of 
\begin{equation}
    \sum_{k=1}^{\infty} \frac{\sin 2\pi kz}{\pi k} = \frac{\sign(z)}{2} - z,
\end{equation}
for $z\in(-1,1)$. Setting $z= F(x_i)-F(x_j)$ and performing the summation for each term
\begin{equation}
    \frac{\psi_T(x_1, x_2, x_3)}{\phi(x_1)\phi(x_2)\phi(x_3)} =\sign\left[F\left(x_1 \right)-F\left(x_2\right)\right]+\sign\left[F\left(x_2\right)-F\left(x_3\right)\right]+\sign\left[F\left(x_3\right)-F\left(x_1\right)\right].
\end{equation}
Applying that $\sign[F(x_i)-F(x_j)]=\sign[x_i-x_j]$, results in 
\begin{equation}
    \frac{\psi_T(x_1, x_2, x_3)}{\phi(x_1)\phi(x_2)\phi(x_3)} =\sign\left(x_1- x_2\right)+\sign\left(x_2-x_3\right)+\sign\left(x_3-x_1\right).
\end{equation}
It is left to prove that, for all possible values of $x_1,x_2,x_3$, $X=Y$, where 
\begin{equation}
    X=\sign\left(x_1- x_2\right)+\sign\left(x_2-x_3\right)-\sign\left(x_1-x_3\right),
\end{equation}
and 
\begin{equation}
    Y=\sign(x_1-x_2)\sign(x_1-x_3)\sign(x_2-x_3). 
\end{equation}
Clearly, $Y$ can only be either $+1$, $-1$ or $0$ if two of the three positions are equivalent. If two of the three positions are equivalent, then, it is also direct to see that $X=0$. It can be seen that $X^3=7X-6Y$. If $Y=1$, $X=1$, $X=2$ and $X=-3$ are all possible solutions. However $X=2$ is not possible and $X=-3$ implies a contradiction of the type $x_1<x_2<x_3$ and  $x_3<x_1<x_2$. If $Y=-1$, $X=-1$, $X=-2$ and $X=3$ are all possible solutions. $X=-2$ is again not possible and $X=3$ implies again a similar contradiction. Therefore, when $Y=0$, $X=0$ and when $Y=\pm1$, $X=\pm1$  meaning that $X=Y$ and, therefore, $\psi_T=\psi_F$. \\ 

\noindent
Another, simpler way of showing that $Y = X$ is to realize that the three signs appearing in $X$ and $Y$ are the same. Importantly, in $X$, $\sign(x_1 - x_3)$ appears with a minus sign in front. To have $Y = 1$, we must have all three signs being positive $(+++)$ or two of them being negative $(+--)$. In the $(+++)$ case, it is clear that $X = 1$. In the $(+--)$ case, $X$ can be $1$ or $-3$ if the positive sign is $\sign(x_1 - x_3)$. However, in the previous paragraph, we have already discussed that $X = -3$ brings a contradiction of the type $x_1 < x_2 < x_3$ and $x_3 < x_1 < x_2$. Therefore, in the $(+--)$ case, $X$ is also $1$. To have $Y = -1$, we must have all three signs being negative $(---)$, which clearly implies $X = -1$, or two of them being positive $(-++)$. Again, in the $(-++)$ case, $X$ can be $-1$ or $3$ if the minus sign is $\sign(x_1 - x_3)$. However, $X = 3$ implies a similar contradiction to the one in the $X = -3$ case. With this reasoning, we can show that $X = Y$ without having to compute and solve an equation containing $X^3$ and $X$.

\boldmath
\section{Proof of $\psi_T=\psi_F$ for the $N$-even case}
\unboldmath
\noindent
The proof is done by induction. We have already proved that $\psi^{(2)}_T(x_1,x_2)=\psi^{(2)}_F(x_1,x_2)$. We will assume that $\psi^{(N-2)}_T(x_1,...,x_{N-2}) = \psi^{(N-2)}_F(x_1,...,x_{N-2})$ and prove 
\begin{equation}
    O\equiv \int \psi^{(N)}_T(x_1,...,x_N)\psi^{(N)}_F(x_1,...,x_N)dx_1...dx_N=1,
\end{equation}
which implies that $\psi^{(N)}_T(x_1,...,x_N)=\psi^{(N)}_F(x_1,...,x_N)$.\\
\noindent
To do so we will need to derive a recursive formula for both $\psi_T$ and $\psi_F$. The $\psi_F$ recursive formula is rather direct to obtain, 
\begin{equation}
\psi^{(N)}_F(x_1, ..., x_N) = \phi(x_N)\phi(x_{N-1}) \left( \prod_{j=1}^{N-1} \sign(x_j - x_N) \right) \left( \prod_{k=1}^{N-2} \sign(x_{k} - x_{N-1}) \right) \times \psi^{(N-2)}_F(x_1, ..., x_{N-2}).
\end{equation}
Of course, one can choose the tag of the two added particles. If the two added particles are in positions $x_i$, $x_j$ where $i<j$, 
\begin{equation}
\label{recursive F}
\begin{aligned}
\psi^{(N)}_F(x_1, ..., x_N) &= \phi(x_i)\phi(x_j) (-1)^{i-1}\left( \prod_{l \neq i}^{N} \sign(x_i - x_l) \right) (-1)^{j-2}\left( \prod_{k \neq j, i}^{N} \sign(x_j - x_k) \right) \\
&\quad \times \psi^{(N-2)}_F(x_1, ..., x_{i-1}, x_{i+1}, ..., x_{j-1}, x_{j+1}, ..., x_N),
\end{aligned}
\end{equation}
where the $(-1)^{i-1}$ and $(-1)^{j-2}$ account for the necessary sign changes in order to always have $\sign(x_l-x_k)$ where $l<k$. \\
\noindent
Next, we derive a recursive formula for the proposed $\psi_T$,
\begin{equation}
    \psi_T^{(N)}(x_1,...,x_N)=\sum_{k_1<...<k_{N_p}}\sqrt{p_N(k_1,...,k_{N_p})}\varphi^{(N)}_{k_1,...,k_{N_p}}(x_1,...,x_N),
\end{equation}
where $\varphi^{(N)}_{k_1,...,k_{N_p}}(x_1,...,x_N)$ is a Slater determinant containing the $N$ states $\chi_{k_1+}$, $\chi_{k_1-}$, ..., $\chi_{k_{N_P}+}$, $\chi_{k_{N_P}-}$ and $N_P=N/2$. Decomposing the Slater determinant in two states minors we get to
\begin{equation}
\begin{aligned}
\psi_T^{(N)}(x_1, \ldots, x_N) =& \!\!\!\! \!\!\sum_{k_1 <...<k_{N_p}}\!\!\!\!\!\! \sqrt{p(k_1, \ldots, k_{N_p})} \frac{\sqrt{(N-2)!}}{\sqrt{N!}} \Bigg[
\det \begin{pmatrix}
\chi_{k_1+}(x_1) & \chi_{k_1-}(x_1) \\
\chi_{k_1+}(x_2) & \chi_{k_1-}(x_2)
\end{pmatrix} \varphi^{(N-2)}_{k_2 \ldots k_{N_p}}(x_3, \ldots ,x_N) \\
-&\det \begin{pmatrix}
\chi_{k_1+}(x_1) & \chi_{k_1-}(x_1) \\
\chi_{k_1+}(x_3) & \chi_{k_1-}(x_3)
\end{pmatrix} \varphi^{(N-2)}_{k_2 \ldots k_{N_p}}(x_2, x_4,\ldots, x_N)\\
+&\det \begin{pmatrix}
\chi_{k_1+}(x_1) & \chi_{k_1-}(x_1) \\
\chi_{k_1+}(x_4) & \chi_{k_1-}(x_4)
\end{pmatrix} \varphi^{(N-2)}_{k_2 \ldots k_{N_p}}(x_2, x_3, x_5, \ldots, x_N)\\ 
&+\ldots
\Bigg],
\end{aligned}
\end{equation}
where the sign preceding the minor $x_i,x_j$ is $+1$ if $i+j$ is odd and $-1$ if $i+j$ is even. The $\sqrt{(N-2)!}$ comes from expressing the purely mathematical determinant of $N-2$ particles as a properly normalised fermionic state. The two-minors determinants can be written as 
\begin{equation}
\label{eq: proper det}
    \det \begin{pmatrix}
\chi_{k+}(x_a) & \chi_{k-}(x_a) \\
\chi_{k+}(x_b) & \chi_{k-}(x_b)
\end{pmatrix}
= 2\phi(x_a)\phi(x_b) \sin[(2k-1)\pi(y_a - y_b)], 
\end{equation} 
where $y=F(x)$. Then, 
\begin{equation}
\begin{aligned}
\psi_T^{(N)}(x_1,..., x_N) = &\!\!\!\! \sum_{k_1 < ...<k_{N_p}} \!\!\!\!\sqrt{p(k_1, \ldots, k_{N_p})} \frac{\sqrt{(N-2)!}}{\sqrt{N!}} \times \\
&\Bigg[2\phi(x_1)\phi(x_2)
\sin{[(2k_1-1)\pi(y_1-y_2)]} \varphi^{(N-2)}_{k_2 \ldots k_N}(x_3, \ldots, x_N) \\
&-2\phi(x_1)\phi(x_3)\sin{[(2k_1-1)\pi(y_1-y_3)]} \varphi^{(N-2)}_{k_2 \ldots k_N}(x_2, x_4, \ldots ,x_N)\\ 
&+2\phi(x_1)\phi(x_4)\sin{[(2k_1-1)\pi(y_1-y_4)]} \varphi^{(N-2)}_{k_2 \ldots k_N}(x_2, x_3,x_5, \ldots ,x_N)+...
\Bigg]. 
\end{aligned}
\end{equation}
The summation can be simplified using that, 
\begin{equation}
    \sum_{k_1 < \ldots<k_{N_p}} =\frac{1}{N_p!}\sum_{k_1\neq \ldots\neq k_{N_P}}=\frac{1}{N_p!}\sum_{k_1,\ldots,k_{N_P}},
\end{equation}
where the last equality is only valid for the specific function we are considering. This is because terms with repeated $k's$ evaluate to zero since a Slater determinant with a repeated column evaluates to zero. Then, 
\begin{equation}
\begin{aligned}
\psi_T^{(N)}(x_1, \ldots, x_N) = &\frac{\sqrt{(N-2)!}}{\sqrt{N!}} \frac{1}{(N/2)!}\sum_{k_1,...,k_{N_p}} \sqrt{p(k_1, \ldots, k_{N_p})}  \times \\ 
&\Bigg[2\phi(x_1)\phi(x_2)
\sin{[(2k_1-1)\pi(y_1-y_2)]} \varphi^{(N-2)}_{k_2 \ldots k_N}(x_3, \ldots, x_N) \\
&-2\phi(x_1)\phi(x_3)\sin{[(2k_1-1)\pi(y_1-y_3)]} \varphi^{(N-2)}_{k_2 \ldots k_N}(x_2, x_4, \ldots ,x_N)\\ 
&+2\phi(x_1)\phi(x_4)\sin{[(2k_1-1)\pi(y_1-y_4)]} \varphi^{(N-2)}_{k_2 \ldots k_N}(x_2, x_3,x_5, \ldots ,x_N)+...
\Bigg]. 
\end{aligned}
\end{equation}
Explicitly writing $p(k_1,...,k_{N_p})$

\begin{equation}
\begin{aligned}
\psi_T^{(N)}(x_1,...,x_N) =& \frac{\sqrt{(N-2)!}}{\sqrt{N!}} \frac{1}{(N/2)!}\frac{\sqrt{N!}}{\pi^{N/2}}2^{N/2}\sum_{k_1,...,k_{N_p}} \frac{1}{(2k_1-1)}\times  ... \times \frac{1}{(2k_{Np}-1)} \times\\ 
&\Bigg[2\phi(x_1)\phi(x_2)
\sin{[(2k_1-1)\pi(y_1-y_2)]} \varphi^{(N-2)}_{k_2 \ldots k_N}(x_3, \ldots, x_N) \\
&-2\phi(x_1)\phi(x_3)\sin{[(2k_1-1)\pi(y_1-y_3)]} \varphi^{(N-2)}_{k_2 \ldots k_N}(x_2, x_4, \ldots ,x_N)\\ 
&+2\phi(x_1)\phi(x_4)\sin{[(2k_1-1)\pi(y_1-y_4)]} \varphi^{(N-2)}_{k_2 \ldots k_N}(x_2, x_3,x_5, \ldots ,x_N)+...
\Bigg]. 
\end{aligned}
\end{equation}
Since, $k_1$ is completely factorised from other $k's$ we can perform the summation over $k_1$. To do so we apply 
\begin{equation}
    \label{eq: proper sum}
    \sum_{k_1} \frac{\sin[(2k_1 - 1)\pi (y_a - y_b)]}{2k_1 - 1} = \frac{\pi}{4} \sign(y_a - y_b),
\end{equation}
for $(y_a - y_b)\in(-1,1)$. Then, 
\begin{equation}
\begin{aligned}
\psi_T^{(N)}(x_1,..., x_N) =& \frac{\sqrt{(N-2)!}}{\sqrt{N!}} \frac{1}{(N/2)!}\frac{\sqrt{N!}}{\pi^{N/2}}2^{N/2}\frac{\pi}{2}\sum_{k_2,...k_{N_p}} \frac{1}{(2k_2-1)}\times  ... \times \frac{1}{(2k_{Np}-1)} \times\\ 
&\Bigg[\phi(x_1)\phi(x_2)
\sign(y_1-y_2)] \varphi^{(N-2)}_{k_2 \ldots k_N}(x_3, \ldots, x_N) \\
&-\phi(x_1)\phi(x_3)\sign(y_1-y_3)] \varphi^{(N-2)}_{k_2 \ldots k_N}(x_2, x_4, \ldots ,x_N)\\ 
&+\phi(x_1)\phi(x_4)\sign(y_1-y_4)]\varphi^{(N-2)}_{k_2 \ldots k_N}(x_2, x_3,x_5, \ldots ,x_N)+...
\Bigg]. 
\end{aligned}
\end{equation}
Regrouping prefactors and changing back the summation to $k_2<...<k_{N_p}$, 
\begin{equation}
\begin{aligned}
\psi_T^{(N)}(x_1,..., x_N) =&  \frac{\sqrt{(N-2)!}}{N/2}\frac{1}{\pi^{(N-2)/2}}\sum_{k_2<...<k_{N_p}} \frac{1}{(k_2-1/2)}\times  ... \times \frac{1}{(k_{Np}-1/2)} \times\\ 
&\Bigg[\phi(x_1)\phi(x_2)
\sign(y_1-y_2)] \varphi^{(N-2)}_{k_2 \ldots k_N}(x_3, \ldots, x_N) \\
&-\phi(x_1)\phi(x_3)\sign(y_1-y_3)] \varphi^{(N-2)}_{k_2 \ldots k_N}(x_2, x_4, \ldots ,x_N)\\ 
&+\phi(x_1)\phi(x_4)\sign(y_1-y_4)]\varphi^{(N-2)}_{k_2 \ldots k_N}(x_2, x_3,x_5, \ldots ,x_N)+...
\Bigg]. 
\end{aligned}
\end{equation}
Identifying $p(k_2,...,k_{N_p})$, and applying $\sign(y_i-y_j)=\sign(x_i-x_j)$ the wave function can be written as a function of the hypothesised wave function for $N-2$ particles: 
\begin{equation}
\begin{aligned}
\label{recursive formula}
\psi_T^{(N)}(x_1 \ldots x_N) =& \frac{2}{N}\Bigg[\phi(x_1)\phi(x_2)\sign(x_1-x_2)\psi^{(N-2)}_T(x_3, ..., x_N)\\ 
&-\phi(x_1)\phi(x_3)\sign(x_1-x_3)\psi^{(N-2)}_T(x_2,x_4, ..., x_N) \\ 
&+\phi(x_1)\phi(x_4)\sign(x_1-x_4)\psi^{(N-2)}_T(x_2,x_3,x_5,..., x_N) ... \Bigg] .
\end{aligned}
\end{equation}
Where the sum goes over all the minors in the initial slater determinant. Now that we have deduced a recursive formula for both $\psi_F$ and $\psi_T$ we proceed to prove that $O=1$ by induction.  We start with 
\begin{equation}
\label{3} 
\begin{aligned}
O=\int&\psi^{(N)}_T(x_1,...,x_N)\psi^{(N)}_F(x_1,...,x_N)dx_1...dx_N=\\ 
    &\frac{2}{N}\Bigg[\int \phi(x_1)\phi(x_2)\sign(x_1-x_2)\psi_T^{(N-2)}(x_3, ..., x_N)\psi^{(N)}_F(x_1,...,x_N) dx_1...dx_N\\ 
   &-\int \phi(x_1)\phi(x_3)\sign(x_1-x_3)\psi_T^{(N-2)}(x_2,x_4, ..., x_N)\psi^{(N)}_F(x_1,...,x_N) dx_1...dx_N\\ 
    +&\int \phi(x_1)\phi(x_4)\sign(x_1-x_4)\psi_T^{(N-2)}(x_2,x_3,x_5 ..., x_N)\psi^{(N)}_F(x_1,...,x_N) dx_1...dx_N +...\Bigg] .
\end{aligned}
\end{equation}
The sign before each term $x_i,x_j$ can be generalised as $(-1)^{i+j+1}$. With this we focus on a given term $x_i,x_j$ which we will denote $O_{ij}$, 
\begin{equation}
\begin{aligned}
O_{ij} =  (-1)^{i+j+1} \int \phi(x_i)\phi(x_j)\sign(x_i-x_j) \psi_T^{(N-2)}(x_1, ..., x_{i-1}, x_{i+1}, ..., x_{j-1}, x_{j+1}, ..., x_N) \psi^{(N)}_F(x_1,...,x_N) dx_1 \ldots dx_N.
\end{aligned}
\end{equation}
To simplify the calculus we make the following change of variables $y=F(x)=\int \phi(x)^2 dx$, $dy=\phi(x)^2 dx $, and apply that $\sign(x_i-x_j)=\sign(y_i-y_j)$. 
Also, we define
\begin{equation}
    \bar{\psi}^{(N)}(x_1,...,x_N)=\frac{\psi^{(N)}(x_1,...,x_N)}{\phi(x_1)\times...\times\phi(x_N)}.
\end{equation}
Applying both the change of notation and change of variables we get to 
\begin{equation}
\label{2}
\begin{aligned}
    O_{ij} =  (-1)^{i+j+1} \int_0^1 \sign(y_i-y_j)\bar{\psi}_T^{(N-2)}(y_1, ..., y_{i-1}, y_{i+1}, ..., y_{j-1}, y_{j+1}, ..., y_N) \bar{\psi}^{(N)}_F(y_1, ..., y_N) \, dy_1 \ldots dy_N.
\end{aligned}
\end{equation}
Next, we rewrite $\bar{\psi}^{(N)}_F(y_1,...,y_N)$ using the recursive form such that the two particles added are $x_i$ and $x_j$. This is 
\begin{equation}
\label{1}
\begin{aligned}
\bar{\psi}^{(N)}_F(y_1, ..., y_N) &= (-1)^{i-1}\left( \prod_{l \neq i}^{N} \sign(y_i - y_l) \right) (-1)^{j-2}\left( \prod_{k \neq j, i}^{N} \sign(y_j - y_k) \right) \\
&\quad \times \bar{\psi^{(N-2)}_F(y_1, ..., y_{i-1}, y_{i+1}, ..., y_{j-1}, y_{j+1}, ..., y_N)}. 
\end{aligned}
\end{equation}
When inserting Eq.~(\ref{1}) into Eq.~(\ref{2}) and applying that for the $N-2$ case $\psi_T^{(N-2)}=\psi_F^{(N-2)}$ (induction), 
\begin{equation}
\begin{aligned}
    \bar{\psi}^{(N-2)}_F(y_1, \ldots, y_{i-1}, y_{i+1}, \ldots, y_{j-1}, y_{j+1}, \ldots, y_N)  \bar{\psi}^{(N-2)}_T(y_1, \ldots, y_{i-1}, y_{i+1}, \ldots, y_{j-1}, y_{j+1}, \ldots, y_N) \\
    = \bar{\psi}^{(N-2)}_F(y_1, \ldots, y_{i-1}, y_{i+1}, \ldots, y_{j-1}, y_{j+1}, \ldots, y_N)^2 = 1,
\end{aligned}
\end{equation}
where $\bar{\psi}_F^{(N-2)}(y_1,..,y_N)^2=1$ since  $\bar{\psi}_F$ is a product of signs, we get to 
\begin{equation}
    O_{ij}=(-1)^{2(i+j)-2}\int_0^1 \sign(y_i-y_j)\left( \prod_{l \neq i}^{N} \sign(y_i - y_l) \right) \left( \prod_{k \neq j, i}^{N} \sign(y_j - y_k) \right) dy_1...dy_N. 
\end{equation}
where $(-1)^{2(i+j)-2}=1$ for any $i,j$. Then, simplifying $\sign(y_i-y_j)$ we get to 
\begin{equation}
    O_{ij}=\int_0^1 \left( \prod_{l \neq i,j}^{N} \sign(y_i - y_l)  \sign(y_j - y_l) \right) dy_1...dy_N, 
\end{equation}
where now it is clear that the choice of $i$ and $j$ is arbitrary and all terms in Eq.~(\ref{3}) are completely equivalent. It is easy to check that there are $\frac{N(N-1)}{2}$ terms since this is the number of 2-states minors in the initial slater determinant. Therefore, the initial integral is greatly simplified to 
\begin{equation}
\begin{aligned}
O=(N-1) \int_0^1 \left( \prod_{l \neq 1,2}^{N} \sign(y_1 - y_l)  \sign(y_2 - y_l) \right) dy_1...dy_N , 
\end{aligned}
\end{equation}
where we have arbitrarily decided that $i,j=1,2$. One can see that the $N-2$ integrals over $y_3,...,y_N$ are equivalent and independent among themselves, thus 
\begin{equation}
\begin{aligned}
O=&(N-1) \int_0^1\int_0^1 \Bigg(\int_0^1 \sign(y_1 - y_3)  \sign(y_2 - y_3) dy_3\Bigg)^{N-2} dy_1dy_2\\ 
=& (N-1)\int_0^1\int_0^1 \Bigg(2(y_1-y_2)\sign(y_2-y_1)+1\Bigg)^{N-2} dy_1dy_2=1, 
\end{aligned}
\end{equation}
when $N$ is even. 

\boldmath
\section{Proof of $\psi_T=\psi_F$ for the $N$-odd case}
\unboldmath
\noindent
The proof is done again by induction.  We have already proved that $\psi^{(3)}_T(x_1,x_2,x_3)=\psi^{(3)}_F(x_1,x_2,x_3)$. We will follow the same procedure as in the even case and assume that $\psi^{(N-2)}_T(x_1,...,x_{N-2})=\psi^{(N-2)}_F(x_1,...,x_{N-2})$ to prove 
\begin{equation}
    O\equiv \int \psi^{(N)}_T(x_1,...,x_N)\psi^{(N)}_F(x_1,...,x_N)dx_1...dx_N=1,
\end{equation}
which implies that $\psi^{(N)}_T(x_1,...,x_N)=\psi^{(N)}_F(x_1,...,x_N)$.
The  $\psi_F$ recursive formula derived in Eq.~(\ref{recursive F}) for the even case is also valid in the odd case. The case of $\psi_T$ is not that simple. Again, we start with our wave function, now for the odd case,
\begin{equation}
    \psi_T^{(N)}(x_1,...,x_N)=\sum_{k_1<...<k_{N_p}}\sqrt{p(k_1,...,k_{N_p})}\varphi^{(N)}_{0,k_1,...,k_{N_p}}(x_1,...,x_N),
\end{equation}
where $\varphi^{(N)}_{0,k_1,...,k_{N_p}}(x_1,...,x_N)$ is the Slater determinant containing the $N$ states $\chi_{k_1+}$, $\chi_{k_1-}$, ..., $\chi_{k_{N_P}+}$, $\chi_{k_{N_P}-}$, and importantly, $\phi$. The number of pairs in the odd case is $N_P=(N-1)/2$. Decomposing the Slater determinant in two states minors we get to
\begin{equation}
\begin{aligned}
\psi_T^{(N)}(x_1, \ldots, x_N) = &\!\!\! \sum_{k_1 <...<k_{N_p}} \sqrt{p(k_1, \ldots, k_{N_p})}\frac{\sqrt{(N-2)!}}{\sqrt{N!}}  \times \\ 
&\Bigg[
\det \begin{pmatrix}
\chi_{k_1+}(x_1) & \chi_{k_1-}(x_1) \\
\chi_{k_1+}(x_2) & \chi_{k_1-}(x_2)
\end{pmatrix} \varphi^{(N-2)}_{0,k_2 \ldots k_{N_p}}(x_3, \ldots ,x_N) \\
-&\det \begin{pmatrix}
\chi_{k_1+}(x_1) & \chi_{k_1-}(x_1) \\
\chi_{k_1+}(x_3) & \chi_{k_1-}(x_3)
\end{pmatrix} \varphi^{(N-2)}_{0,k_2 \ldots k_{N_p}}(x_2, x_4,\ldots, x_N)\\
+&\det \begin{pmatrix}
\chi_{k_1+}(x_1) & \chi_{k_1-}(x_1) \\
\chi_{k_1+}(x_4) & \chi_{k_1-}(x_4)
\end{pmatrix} \varphi^{(N-2)}_{0,k_2 \ldots k_{N_p}}(x_2, x_3, x_5, \ldots, x_N)\\ 
&+\ldots
\Bigg],
\end{aligned}
\end{equation}
where the sign preceding the minor $x_i,x_j$ is $+1$ if $i+j$ is odd and $-1$ if $i+j$ is even. The $\sqrt{(N-2)!}$ comes from expressing the purely mathematical determinant of $N-2$ particles as a properly normalised fermionic state. The two-minors determinants can be now written as
\begin{equation}
\label{eq: proper det odd}
    \det \begin{pmatrix}
\chi_{k+}(x_a) & \chi_{k-}(x_a) \\
\chi_{k+}(x_b) & \chi_{k-}(x_b)
\end{pmatrix}
= 2\phi(x_a)\phi(x_b) \sin[2k\pi(y_a - y_b)],
\end{equation} 
where $y=F(x)$. Then, 

\begin{equation}
\begin{aligned}
\psi_T^{(N)}(x_1,..., x_N) = &\!\!\!\! \sum_{k_1 < ...<k_{N_p}} \!\!\!\!\sqrt{p(k_1, \ldots, k_{N_p})} \frac{\sqrt{(N-2)!}}{\sqrt{N!}} \times \\
&\Bigg[2\phi(x_1)\phi(x_2)
\sin{[2k_1\pi(y_1-y_2)]} \varphi^{(N-2)}_{0,k_2 \ldots k_N}(x_3, \ldots, x_N) \\
&-2\phi(x_1)\phi(x_3)\sin{[2k_1\pi(y_1-y_3)]} \varphi^{(N-2)}_{0,k_2 \ldots k_N}(x_2, x_4, \ldots ,x_N)\\ 
&+2\phi(x_1)\phi(x_4)\sin{[2k_1\pi(y_1-y_4)]} \varphi^{(N-2)}_{0,k_2 \ldots k_N}(x_2, x_3,x_5, \ldots ,x_N)+...
\Bigg]. 
\end{aligned}
\end{equation}
As in the even case, we simplify the summation,
\begin{equation}
\begin{aligned}
\psi_T^{(N)}(x_1, \ldots, x_N) = &\frac{\sqrt{(N-2)!}}{\sqrt{N!}} \frac{1}{[(N-1)/2]!}\sum_{k_1,...,k_{N_p}} \sqrt{p(k_1, \ldots, k_{N_p})}  \times \\ 
&\Bigg[2\phi(x_1)\phi(x_2)
\sin{[2k_1\pi(y_1-y_2)]} \varphi^{(N-2)}_{k_2 \ldots k_N}(x_3, \ldots, x_N) \\
&-2\phi(x_1)\phi(x_3)\sin{[2k_1\pi(y_1-y_3)]} \varphi^{(N-2)}_{k_2 \ldots k_N}(x_2, x_4, \ldots ,x_N)\\ 
&+2\phi(x_1)\phi(x_4)\sin{[2k_1\pi(y_1-y_4)]} \varphi^{(N-2)}_{k_2 \ldots k_N}(x_2, x_3,x_5, \ldots ,x_N)+...
\Bigg]. 
\end{aligned}
\end{equation}
Explicitly writing $p(k_1,...,k_{N_p})$

\begin{equation}
\begin{aligned}
\psi_T^{(N)}(x_1,...,x_N) =& \frac{\sqrt{(N-2)!}}{\sqrt{N!}} \frac{1}{[(N-1)/2]!}\frac{\sqrt{N!}}{\pi^{(N-1)/2}}\sum_{k_1,...,k_{N_p}} \frac{1}{k_1}\times  ... \times \frac{1}{k_{Np}} \times\\ 
&\Bigg[2\phi(x_1)\phi(x_2)
\sin{[2k_1\pi(y_1-y_2)]} \varphi^{(N-2)}_{k_2 \ldots k_N}(x_3, \ldots, x_N) \\
&-2\phi(x_1)\phi(x_3)\sin{[2k_1\pi(y_1-y_3)]} \varphi^{(N-2)}_{k_2 \ldots k_N}(x_2, x_4, \ldots ,x_N)\\ 
&+2\phi(x_1)\phi(x_4)\sin{[2k_1\pi(y_1-y_4)]} \varphi^{(N-2)}_{k_2 \ldots k_N}(x_2, x_3,x_5, \ldots ,x_N)+...
\Bigg]. 
\end{aligned}
\end{equation}
Since $k_1$ is completely factorised from other $k's$ we perform the summation over $k_1$ applying
\begin{equation}
\label{proper sum odd}
    \sum_{n=1}^{\infty} \frac{\sin  [2\pi n(y_a-y_b)]}{\pi n} = \frac{\sign(y_a-y_b)}{2}- (y_a-y_b),
\end{equation}
for $y_a-y_b\in(-1,1)$. Then, 
\begin{equation}
\begin{aligned}
\psi_T^{(N)}(x_1,..., x_N) =&  \frac{\sqrt{(N-2)!}}{\sqrt{N!}} \frac{2}{[(N-1)/2]!}\sqrt{N!}\sum_{k_2,...,k_{N_p}} \frac{1}{\pi k_2}\times  ... \times \frac{1}{\pi k_{Np}}  \times\\ 
&\Bigg[\phi(x_1)\phi(x_2)
\left( \frac{\sign(y_1-y_2)}{2}-(y_1-y_2)\right) \varphi^{(N-2)}_{k_2 \ldots k_N}(x_3, \ldots, x_N) \\
&-\phi(x_1)\phi(x_3)\left( \frac{\sign(y_1-y_3)}{2}-(y_1-y_3)\right)\varphi^{(N-2)}_{k_2 \ldots k_N}(x_2, x_4, \ldots ,x_N)\\ 
&+\phi(x_1)\phi(x_4)\left( \frac{\sign(y_1-y_4)}{2}-(y_1-y_4)\right)\varphi^{(N-2)}_{k_2 \ldots k_N}(x_2, x_3,x_5, \ldots ,x_N)+...
\Bigg]. 
\end{aligned}
\end{equation}
Regrouping prefactors and changing back the summation to $k_2<...<k_{N_p}$,
\begin{equation}
\begin{aligned}
\psi_T^{(N)}(x_1,..., x_N) =&  \sqrt{(N-2)!} \frac{2}{(N-1)/2}\sum_{k_2<...<k_{N_p}} \frac{1}{\pi k_2}\times  ... \times \frac{1}{\pi k_{Np}}  \times\\ 
&\Bigg[\phi(x_1)\phi(x_2)
\left( \frac{\sign(y_1-y_2)}{2}-(y_1-y_2)\right) \varphi^{(N-2)}_{k_2 \ldots k_N}(x_3, \ldots, x_N) \\
&-\phi(x_1)\phi(x_3)\left( \frac{\sign(y_1-y_3)}{2}-(y_1-y_3)\right)\varphi^{(N-2)}_{k_2 \ldots k_N}(x_2, x_4, \ldots ,x_N)\\ 
&+\phi(x_1)\phi(x_4)\left( \frac{\sign(y_1-y_4)}{2}-(y_1-y_4)\right)\varphi^{(N-2)}_{k_2 \ldots k_N}(x_2, x_3,x_5, \ldots ,x_N)+...
\Bigg]. 
\end{aligned}
\end{equation}
Identifying $p(k_2,...,k_{N_p})$, $\psi_T^{(N)}$ can be written as  a function of $\psi_T^{(N-2)}$,
\begin{equation}
\begin{aligned}
\label{recursive formula odd}
\psi_T^{(N)}(x_1 \ldots x_N) =& \frac{4}{N-1}\Bigg[\phi(x_1)\phi(x_2)\left( \frac{\sign(y_1-y_2)}{2}-(y_1-y_2)\right) \psi^{(N-2)}_T(x_3, ..., x_N)\\ 
&-\phi(x_1)\phi(x_3)\left( \frac{\sign(y_1-y_3)}{2}-(y_1-y_3)\right) \psi^{(N-2)}_T(x_2,x_4, ..., x_N) \\ 
&+\phi(x_1)\phi(x_4)\left( \frac{\sign(y_1-y_4)}{2}-(y_1-y_4)\right) \psi^{(N-2)}_T(x_2,x_3,x_5,..., x_N) ... \Bigg] .
\end{aligned}
\end{equation}
Where the sum goes over all the minors in the initial slater determinant. As in the even case, now that we have a recursive formula for both $\psi_F$ and $\psi_T$ we proceed with the actual proof of $O=1$. Applying the recursive formula of $\psi_F$ into the calculation of O,
\begin{equation}
\label{3odd} 
\begin{aligned}
O=&\frac{4}{N-1}\Bigg[\int \phi(x_1)\phi(x_2)\left( \frac{\sign(y_1-y_2)}{2}-(y_1-y_2)\right) \psi_T^{(N-2)}(x_3, ..., x_N)\psi^{(N)}_F(x_1,...,x_N) dx_1...dx_N\\ 
   &-\int \phi(x_1)\phi(x_3)\left( \frac{\sign(y_1-y_3)}{2}-(y_1-y_3)\right) \psi_T^{(N-2)}(x_2,x_4, ..., x_N)\psi^{(N)}_F(x_1,...,x_N) dx_1...dx_N\\ 
    +&\int \phi(x_1)\phi(x_4)\left( \frac{\sign(y_1-y_4)}{2}-(y_1-y_4)\right) \psi_T^{(N-2)}(x_2,x_3,x_5 ..., x_N)\psi^{(N)}_F(x_1,...,x_N) dx_1...dx_N +...\Bigg] ,
\end{aligned}
\end{equation}
where the sign before each term $x_i,x_j$ can be generalised as $(-1)^{i+j+1}$. With this we focus on a given term $x_i,x_j$, 

\begin{multline}
    O_{ij}=(-1)^{i+j+1}\int \phi(x_i)\phi(x_j) \left( \frac{\sign(y_i-y_j)}{2} - (y_i-y_j) \right) \psi_T^{(N-2)}(x_1, ..., x_{i-1}, x_{i+1}, ..., x_{j-1}, x_{j+1}, ..., x_N) \\
    \times \psi^{(N)}_F(x_1, ..., x_N) \, dx_1 \ldots dx_N.
\end{multline}
To simplify the calculus we make the change of variables $y=F(x)=\int \phi(x)^2 dx$, $dy=\phi(x)^2 dx $. Also, we denote by 
\begin{equation}
    \bar{\psi}^{(N)}(x_1,...,x_N)=\frac{\psi^{(N)}(x_1,...,x_N)}{\phi(x_1)\times...\times\phi(x_N)}.
\end{equation}
Applying both the change of notation and change of variables we get to 
\begin{multline}
\label{2 odd}
O_{ij}=(-1)^{i+j+1} \int_0^1 \left( \frac{\sign(y_i-y_j)}{2} - (y_i-y_j) \right) \bar{\psi}_T^{(N-2)}(y_1, ..., y_{i-1}, y_{i+1}, ..., y_{j-1}, y_{j+1}, ..., y_N) \\
\times \bar{\psi}^{(N)}_F(y_1, ..., y_N) dy_1 \ldots dy_N.
\end{multline}
We first focus on the term including the $y_i-y_j$ and forget about the sign. We denote by $I_{ij}$ the following integral, 
\begin{equation}
\label{I}
I_{ij}=(-1)^{i+j} \int_0^1  (y_i-y_j) \bar{\psi}_T^{(N-2)}(y_1, ..., y_{i-1}, y_{i+1}, ..., y_{j-1}, y_{j+1}, ..., y_N)  \bar{\psi}^{(N)}_F(y_1, ..., y_N) dy_1 \ldots dy_N. 
\end{equation}
The integral $I_{ij}$ does not depend on the choice of $i,j$ as long as $i<j$ which occurs in all terms of Eq.~(\ref{3odd}). We rewrite $\bar{\psi}^{(N)}_F(y_1,...,y_N)$ using the recursive form such that the two particles added are $x_i$ and $x_j$. This is 
\begin{equation}
\label{1 odd}
\begin{aligned}
\bar{\psi}^{(N)}_F(y_1, ..., y_N) &= (-1)^{i-1}\left( \prod_{l \neq i}^{N} \sign(y_i - y_l) \right) (-1)^{j-2}\left( \prod_{k \neq j, i}^{N} \sign(y_j - y_k) \right) \\
&\quad \times \bar{\psi}^{(N-2)}_F(y_1, ..., y_{i-1}, y_{i+1}, ..., y_{j-1}, y_{j+1}, ..., y_N). 
\end{aligned}
\end{equation}
When inserting Eq.~(\ref{1}) into Eq.~(\ref{2}) and applying that for the $N-2$ case $\psi_T^{(N-2)}=\psi_F^{(N-2)}$ (induction), 
\begin{equation}
\begin{aligned}
    \bar{\psi}^{(N-2)}_F(y_1, \ldots, y_{i-1}, y_{i+1}, \ldots, y_{j-1}, y_{j+1}, \ldots, y_N)  \bar{\psi}^{(N-2)}_T(y_1, \ldots, y_{i-1}, y_{i+1}, \ldots, y_{j-1}, y_{j+1}, \ldots, y_N) \\
    = \bar{\psi}^{(N-2)}_F(y_1, \ldots, y_{i-1}, y_{i+1}, \ldots, y_{j-1}, y_{j+1}, \ldots, y_N)^2 = 1,
\end{aligned}
\end{equation}
where $\bar{\psi}_F^{(N-2)}(y_1,..,y_N)^2=1$ since  $\bar{\psi}_F$ is a product of signs, we get to 
\begin{equation}
    I_{ij}=(-1)^{2(i+j)-3}\int_0^1 (y_i-y_j)\left( \prod_{l \neq i}^{N} \sign(y_i - y_l) \right) \left( \prod_{k \neq j, i}^{N} \sign(y_j - y_k) \right) dy_1...dy_N. 
\end{equation}
where $(-1)^{2(i+j)-3}=-1$ for any $i,j$. This can be rewritten as,
\begin{equation}
    I_{ij}=-\int_0^1 (y_i-y_j)\sign(y_i-y_j)\left( \prod_{l \neq i,j}^{N} \sign(y_i - y_l)\sign(y_j - y_l) \right) dy_1...dy_N. 
\end{equation}
 For simplicity, we choose $i,j=1,2$ and realise that the $N-2$ integrals over $x_3,...x_N$ are equivalent, then 
\begin{equation}
\begin{aligned}
    I_{ij}=&-\int_0^1\int_0^1 (y_1-y_2)\sign(y_1-y_2)\Bigg( \int_0^1 \sign(y_1 - y_3)\sign(y_2 - y_3) dy_3 \Bigg)^{N-2} dy_1dy_2 \\
    =& -\int_0^1\int_0^1 (y_1-y_2)\sign(y_1-y_2)\Bigg(2(y_1-y_2)\sign(y_2-y_1)+1\Bigg)^{N-2} dy_1dy_2=0 . 
\end{aligned}
\end{equation}
when $N$ is odd. After showing that $I_{ij}=0$, we are left with, 
\begin{equation}
O_{ij}=(-1)^{i+j+1} \int_0^1 \frac{\sign(y_i-y_j)}{2} \bar{\psi}_T^{(N-2)}(y_1, ..., y_{i-1}, y_{i+1}, ..., y_{j-1}, y_{j+1}, ..., y_N) \bar{\psi}^{(N)}_F(y_1, ..., y_N) dy_1 \ldots dy_N.
\end{equation}
Which has a completely equivalent form to the  $O_{ij}$ in the even case shown in Eq.~(\ref{2}). Therefore, we can apply the same procedure and steps already applied in the even case to finally get to 
\begin{equation}
\begin{aligned}
O=& N \int_0^1 \int_0^1 \Bigg( \int_0^1\sign(y_1 - y_3)  \sign(y_2 - y_3)  dy_3\Bigg)^{N-2} dy_1dy_2\\ 
=& N\int_0^1\int_0^1 \Bigg(2(y_1-y_2)\sign(y_2-y_1)+1\Bigg)^{N-2} dy_1dy_2=1, 
\end{aligned}
\end{equation}
when $N$ is odd.

\boldmath
\section{Calculation of $\langle\vec{O}_3^{FTG}\rangle$ }
\unboldmath
\noindent In this section we provide the explicit calculations of the expectation values of 
\begin{equation}
\vec{O}_3^{FTG} = 
\begin{pmatrix}
n_{k+} + n_{k-} + n_{l+} + n_{l-} \\
n_{k+}n_{k-} + n_{l+} n_{l-} \\
a_{k+}^\dagger a_{k-}^\dagger a_{l-} a_{l+} + \text{h.c.}
\end{pmatrix}=
\begin{pmatrix}
2\left(\lambda_k^{(N)}+\lambda_l^{(N)}\right) \\
\lambda_k^{(N)}+\lambda_l^{(N)}\\
2\!\sum_{(k_2 < \ldots < k_{N_P}) \neq l,k} \!\!\sqrt{\!p(l, k_2 \ldots)p(k, k_2 \ldots)}
\end{pmatrix}.
\end{equation}
for the FTG gas case. We start with $\langle n_k\rangle$. 
\begin{equation}
    \langle n_{k+} \rangle =\langle a_{k+}^\dagger a_{k+}\rangle=\langle\psi_{T}|\sum_{k_1<..<k_{N_P}}\sqrt{p_N(k_1,...,k_{Np})}a_{k+}^\dagger a_{k+}P_{k_1}^\dagger ...P_{k_{N_p}}^\dagger |0\rangle =\sum_{(k_2<...<k_{Np})\neq k}p(k,k_2,...,k_{Np})=\lambda_k^{(N)}.
\end{equation}
Of course, the derivation $\langle n_{k-} \rangle=\lambda_k^{(N)}$ is analogous to the latter. 
With this, we get to 
\begin{equation}
    \langle n_{k+}+n_{k-}+n_{l+}+n_{l-}\rangle=2\left(\lambda_k^{(N)}+\lambda_l^{(N)}\right).
\end{equation}
The next expectation value we compute is $\langle n_{k+}n_{k-}\rangle$, 
\begin{equation}
  \langle n_{k+}n_{k-}\rangle=<\psi_{T}|\sum_{k_1<..<k_{N_P}}\sqrt{p_N(k_1,...,k_{Np})}a_{k+}^\dagger a_{k+}a_{k-}^\dagger a_{k-}P_{k_1}^\dagger ...P_{k_{N_p}}^\dagger |0\rangle= \sum_{(k_2<...<k_Np)\neq k}p(k,k_2,...,k_{Np})=\lambda_k^{(N)}.
\end{equation}
With this, we get to 
\begin{equation}
    \langle n_{k+}n_{k-}+ n_{l+}n_{l-} \rangle=\lambda_k^{(N)}+\lambda_l^{(N)}.
\end{equation}
Finally, the last expectation value left to be computed is $\langle a_{k+}^\dagger a_{k-}^\dagger a_{l-} a_{l+}\rangle$. 
\begin{equation}
    \begin{split}
        \langle a_{k+}^\dagger a_{k-}^\dagger a_{l-} a_{l+} \rangle &= \langle \psi_{T} | \sum_{k_1 < \ldots < k_{N_P}} \sqrt{p(k_1, \ldots, k_{N_P})} a_{k+}^\dagger a_{k-}^\dagger a_{l-} a_{l+} P_{k_1}^\dagger ...P_{k_{N_p}}^\dagger |0\rangle  \\
        &= \langle \psi_{T} | \sum_{(k_2 < \ldots < k_{N_P}) \neq l} \sqrt{p(l, k_2, \ldots,k_{N_P})} P_{k}^\dagger ...P_{k_{N_p}}^\dagger |0\rangle\\ 
        &=\sum_{(k_2 < \ldots < k_{N_P}) \neq l,k} \sqrt{p(l, k_2 \ldots,k_{N_P})p(k, k_2, \ldots,k_{N_P})}. 
    \end{split}
\end{equation}
With this, we get to 
\begin{equation}
     \langle a_{k+}^\dagger a_{k-}^\dagger a_{l-} a_{l+} + \text{h.c} \rangle=2\sum_{(k_2 < \ldots < k_{N_P}) \neq l,k} \sqrt{p(l, k_2, \ldots,k_{N_P})p(k, k_2, \ldots,k_{N_P})} .
\end{equation}

\end{document}